\documentclass[twocolumn]{revtex4-2}

\usepackage{dcolumn}
\usepackage{bm}
\usepackage{hyperref}

\usepackage{graphicx}
\usepackage{amsmath}
\usepackage{amssymb}
\usepackage{amsthm}
\usepackage{cleveref}
\usepackage{bbold}
\usepackage[ruled,lined]{algorithm2e}

\newcommand{\be}{\begin{equation}}
\newcommand{\ee}{\end{equation}}
\newcommand{\ba}{\begin{array}}
\newcommand{\ea}{\end{array}}
\newcommand{\bea}{\begin{eqnarray}}
\newcommand{\eea}{\end{eqnarray}}

\DeclareMathOperator{\tr}{Tr}
\newcommand{\ra}{\rangle}
\newcommand{\la}{\langle}
\newcommand{\vc}{\vec c}
\renewcommand{\oc}{c^{\mathrm{Fr}}}

\newcommand{\calH}{{\cal H }}
\newcommand{\calL}{{\cal L }}

\newcommand{\calG}{{\cal G }}

\newcommand{\calC}{{\cal C }}
\newcommand{\calS}{{\cal S }}
\newcommand{\calO}{{\cal O }}

\newcommand{\calU}{{\cal U }}

\newcommand{\RR}{\mathbb{R}}

\newcommand{\comm}{\mathsf{comm}}
\newcommand{\ad}{\mathrm{Ad}}
\newcommand{\exact}{\mathsf{exact}}
\newcommand{\apx}{\mathsf{approx}}
\renewcommand{\comm}{\textbf{\em comm}}

\newtheorem{prop}{Proposition}
\newtheorem{lemma}{Lemma}
\newtheorem{fact}{Fact}

\newtheorem{theorem}{Theorem}

\begin{document}

\preprint{APS/123-QED}

\title{Trotter error bounds and dynamic multi-product formulas for Hamiltonian simulation}

\author{Sergiy Zhuk}
 \email{sergiy.zhuk@ie.ibm.com}
 \affiliation{IBM Quantum, IBM Research Europe - Dublin, IBM Technology Campus, Dublin 15, Ireland}

 \author{Sergey Bravyi}%
 \email{sbravyi@us.ibm.com}
\affiliation{IBM Quantum, IBM T.J. Watson Research Center}
\author{Niall F. Robertson}%
 \email{niall.robertson@ibm.com}
\affiliation{IBM Quantum, IBM Research Europe - Dublin, IBM Technology Campus, Dublin 15, Ireland}

\begin{abstract}
Multi-product formulas (MPF) are linear combinations of Trotter circuits offering
high-quality simulation of Hamiltonian time evolution with fewer Trotter steps.
Here we report two contributions aimed at making multi-product formulas
more viable for near-term quantum simulations.
First, we extend the theory of Trotter error with commutator scaling
developed by Childs, Su, Tran et al. to multi-product formulas.
Our result implies that multi-product formulas can achieve a quadratic reduction of Trotter error in 1-norm (nuclear norm) on arbitrary time intervals
compared with the regular product formulas
without increasing the required circuit depth or qubit connectivity.
The number of circuit repetitions grows only by a constant factor.
Second, we introduce dynamic multi-product formulas with time-dependent coefficients chosen to minimize a certain efficiently computable proxy for the Trotter error. We use a minimax estimation method to make dynamic multi-product formulas robust to uncertainty from algorithmic errors, sampling and hardware noise. We call this method Minimax MPF and we provide a rigorous bound on its error.
\end{abstract}

\maketitle

\section{Introduction}
\label{sec:intro}

Simulation of Hamiltonian dynamics is one of the most natural
use cases for quantum computers.
Such simulation can aid researchers
in computing electronic structure of large molecules~\cite{wang2008quantum,reiher2017elucidating},
finding energy spectrum of elementary excitations~\cite{harle2022observing},
understanding
thermalization in closed systems~\cite{kaufman2016quantum},
and testing theoretical models of a quantum chaos~\cite{garttner2017measuring}.
It is strongly believed that Hamiltonian dynamics simulation is intractable
for  conventional classical computers due to an exponentially large dimension of the Hilbert space
and highly entangled nature of  time-evolved states.
In contrast, quantum computers can efficiently simulate Hamiltonian dynamics for any
spin or fermion system with few-body interactions~\cite{lloyd1996universal} or, more generally, for any sparse Hamiltonian~\cite{low2017optimal}.

Here we focus on a particular class of quantum simulation algorithms based on product formulas.
A product formula is a quantum circuit $\calS(t)$ approximating
the evolution operator  $e^{-itH}$
for a quantum  system with a Hamiltonian $H$ evolved over time $t$.
The circuit $\calS(t)$ is usually chosen as a product of local evolution operators
associated with few-body interactions describing the system.
As such, product formulas inherit local structure of the underlying system
which often obviates the need for long-range entangling gates.
For example, simulating a spin chain Hamiltonian with short-range interactions
would only require a hardware with the linear qubit connectivity.
Furthermore, local structure of product formulas results in improved
approximation error bounds that can exploit commutativity of non-overlapping Hamiltonian terms~\cite{childs2021theory}.
Most of product formulas also exhibit a periodicity
structure such that a few layers of gates simulating
a short time step are repeated many times to simulate a longer evolution time.
The periodicity structure aids error mitigation methods,
such as the Probabilistic Error Cancellation~\cite{temme2017error,berg2022probabilistic},
as one has to learn noise models only for a few distinct layers of  gates.

High-quality product formulas that accurately approximate Hamiltonian dynamics
of practically relevant systems
can be very deep. A common approach is to break the  desired evolution time $t$
into $k$ intervals of length $t/k$ and apply a product formula $k$
times with the evolution time $t/k$.
This yields a circuit $\calS(t/k)^k$ whose depth scales linearly with $k$.
The required  number of time steps $k$ depends on
details of the Hamiltonian,
 desired approximation error,
and the type of a product formula.
For example, solving a benchmark simulation problem posed by Childs, Maslov, et al.~\cite{childs2018toward}
for a system of $100$ qubits would require $k\ge 5000$ time steps
if one uses the fourth-order Trotter  product formula, see Section~F.3 of~\cite{childs2018toward}.
The resulting circuit $\calS(t/k)^k$ would be too deep to execute reliably on  near-term quantum processors
lacking error correction.

Multi-product formulas introduced in the context of quantum computing
by Childs and Wiebe~\cite{childs2012hamiltonian}
and developed further in~\cite{low2019well,vazquez2022well,rendon2022improved}
may provide a more viable path to near-term quantum simulations.
The simplest example of
a multi-product formula (MPF) is a linear combination $\sum_{j=1}^r c_j \calS(t/k_j)^{k_j}$,
where $c_j$ are real or complex coefficients, $\calS(t)$ is some base product formula,
and $k_1,\ldots,k_r$ is a sequence of integers.
The $j$-th term in the MPF approximates the desired  evolution operator $e^{-itH}$ by
performing $k_j$ steps of the base product formula with the evolution time $t/k_j$.
The coefficients $c_j$  can be chosen such that the  errors introduced by each
circuit $\calS(t/k_j)^{k_j}$
approximately  cancel each other enabling high-accuracy simulations with
fewer time steps~\cite{low2019well}.

Refs.~\cite{childs2012hamiltonian,low2019well}
envisioned implementing the whole MPF  on a quantum computer
using the  linear combination of unitaries (LCU) method.
This yields a simulation algorithm with a favorable asymptotic cost that
scales nearly linearly with the evolution time and poly-logarithmically with the desired approximation
error~\cite{low2019well}. However, this algorithm has not been experimentally
demonstrated yet due to a high complexity of LCU circuits.

Here we adopt a simpler implementation of MPFs
due to Vazquez, Egger, et al.~\cite{vazquez2022well},
Rendon, Watkins, and Wiebe~\cite{rendon2022improved},
see also~\cite{endo2019mitigating}.
These authors realized that implementing the whole MPF on a quantum processor
is not needed for certain tasks, such as computing time-evolved expected values of observables~\cite{vazquez2022well}
or quantum phase estimation~\cite{rendon2022improved}.
Instead, these tasks can be accomplished by
implementing each  individual circuit $\calS(t/k_j)^{k_j}$ in the MPF
on a quantum device
 and  classical post-processing of the measured data.
 Such quantum-classical MPF
can be described as a linear combination of density matrices
$\mu(t)=\sum_{j=1}^r c_j \rho_{k_j}(t)$, where $\rho_{k}(t)$ is a state
obtained by applying $k$ steps of the base product formula to the initial
state $\rho_{in}$ at time $t=0$. Equivalently, $\rho_k(t)=\calS(t/k)^k \rho_{in} \calS(t/k)^{-k}$.
Assuming that $\mu(t)$ is a good approximation of the exact time-evolved
state $\rho(t)=e^{-itH} \rho_{in} e^{itH}$, the expected value
of any observable  $\mathrm{Tr}(\calO \rho(t))$ can be approximated
by a linear combination $\sum_{j=1}^r c_j  \mathrm{Tr}(\calO \rho_{k_j}(t))$.
The latter can be computed classically as long as a quantum processor
provides an estimate of the expected values $\mathrm{Tr}(\calO \rho_{k_j}(t))$.

To reap the benefits of MPFs one needs to address two points.
First, one needs to bound the approximation error
$\|\mu(t)-\rho(t)\|_1$. To the best of our knowledge, such bounds
were previously known
only in the special case when
the number of time steps $k_j$
is sufficiently large
so that $e^{-itH/k_j}\approx I$
for all $j$,
see ~\cite{childs2012hamiltonian,rendon2022improved}.
Here we overcome this limitation
by extending the theory of Trotter error
with commutator scaling due to Childs, Su, Tran et al~\cite{childs2021theory} from regular product formulas to MPFs.
We show that an MPF can achieve a quadratic reduction
of the approximation error compared with the base product formula
without increasing the requited circuit depth or required qubit connectivity,
see Theorem~\ref{thm:MPF} in Section~\ref{sec:MPF} for a formal statement.
To the best of our knowledge, this  provides the first
rigorous upper bound on the approximation error achieved by MPFs
which holds for any evolution time $t$ and any number of
time steps $k_j$. Numerical simulations suggest that the error scaling
predicted by  Theorem~\ref{thm:MPF} is nearly tight, see Section~\ref{sec:numerics}.
Our main technical innovation is an integral representation of MPFs
based on
Euler–Maclaurin formula. It allows us  to express
the difference $\mu(t)-\rho(t)$ in terms of nested commutators of the type
studied in~\cite{childs2021theory}.

Secondly, one  needs an efficient method
for  computing MPF coefficients $c_j$. Indeed, the best scenario would have been to find optimal coefficients
minimizing the error $\|\mu(t)-\rho(t)\|_1$ at any given time $t$, which is equivalent to projecting $\rho$ onto MPF
subspace, namely the subspace generated by $\rho_{k_1}\dots\rho_{k_r}$. However, in practice this may not be possible since the exact solution $\rho(t)$ is unknown. Instead, a common strategy is the polynomial extrapolation method~\cite{chin2010multi}.
It works by expanding $\rho_{k_j}(t)$
in powers of $1/k_j$. This gives
Taylor  series with the $0$-th order term $\rho(t)$ corresponding to the limit $k_j\to\infty$
and higher order terms proportional to positive powers of $1/k_j$.
By choosing the coefficients $c_j$ as a solution of a suitable linear system
one can ensure that all unwanted lower-order terms in the above expansion  cancel each other.
This is the approach taken in~\cite{childs2012hamiltonian,low2019well,vazquez2022well,rendon2022improved}
as well as in our upper bound on the MPF approximation error.
However, the polynomial extrapolation method is sub-optimal as it forces the coefficients $c_j$ to be time independent
and ignores all structure of the simulated system.

Another important practical issue one needs to deal with while using MPF is that the coefficents of MPF formula are sensitive to the choice of $k_i$ as for certain $k_i$ the resulting Vandermonde matrix~\eqref{ZNE} is ill-conditioned numerically, hence the resulting vector of coefficients has large condition number $\sum_i |c_i|$  and so $c_i$ might amplify sampling noise when computing the linear combination $\sum_{i=1}^r c_i \tr(O\rho_{k_i}(t))$ provided $\tr(O\rho_{k_i}(t))$ is computed on a noisy quantum hardware. Restricting $k_i$ one may obtain so called well-conditioned MPFs~\cite{vazquez2022well}.

To overcome the aforementioned limitations 
we propose dynamic multi-product formulas.
These are MPFs of the form
$\mu^D(t)=\sum_{j=1}^r c_j(t) \rho_{k_j}(t)$
with time-dependent coefficients $c_j(t)$
chosen to minimize the error $\|\mu^D(t)-\rho(t)\|_F$ measured in Frobenious norm. Dynamic MPF with optimal coefficients -- let's denote them by $\oc(t)$, represents optimal projection of $\rho(t)$ onto MPF subspace w.r.t. Frobenious inner product. It will be demonstrated later (and in fact it is quite easy to see) that the optimal coefficients $\oc(t)$ require computing overlaps between
Trotter circuits such as $\mathrm{Tr}(\rho_{k_i}(t) \rho_{k_j}(t))$ and also overlaps of $\rho(t)$ with circuits such as $\mathrm{Tr}(\rho(t) \rho_{k_j}(t))$. The former can be efficiently estimated (with a small additive error) on a quantum computer (see~\Cref{sec:DMPF} for details). But overlaps such as $\mathrm{Tr}(\rho(t) \rho_{k_j}(t))$ require extra
considerations since $\rho(t)$ is not available.
To approximate $\rho(t)$, let $dt>0$ be a sufficiently small time step. Assume without the loss of generality that $\mu^D(t-dt)$ is known and is a good proxy for $\rho(t-dt)$ (this is the case at least initially!). We approximate the unknown $\rho(t)$ by $\calS((dt)/k_0)^{k_0} \mu^D(t-dt)\calS((dt)/k_0)^{-k_0}$ for small $dt>0$ and a natural $k_0$. Simple linear algebra allows us to formalize this process in the form of a linear discrete-time dynamical system which describes dynamics of the optimal projection coefficients $\oc(t)$ over time. This system is subject to uncertainty: namely, the approximation error of
$\rho(t)$ by $\mu^D(t-dt)$, and the sampling error in estimating $\mathrm{Tr}(\rho_{k_i}(t) \rho_{k_j}(t))$ on the device. To deal with this uncertainty we employ minimax estimation method~\cite{zhuk2010minimax}, an estimation technique designed to handle uncertainty in dynamical systems \cite{zhuk2021minimax,zhuk2023detectability}, to find an estimate of the vector of optimal coefficients  $\oc$ in the presence of noise. We prove that this estimate is robust to uncertainty due to algorithmic errors and sampling noise. We refer to this technique as minimax MPF. A practically important advantage of minimax MPF over the static MPF is that the former can provide well-conditioned coefficients $c_i$ (i.e. $\sum_i |c_i|$ is small) for any sequence of $k_i$ in contrast to the static MPF which is well-conditioned only for a restricted set of $k_i$, and to find the latter set one needs to solve a combinatorial optimization problem
(see Figure~\ref{fig:table_examples} below). The reason that minimax MPF provides well-conditioned coefficients is that the method involves a penalization term which is the product of the magnitude of the sampling error and the norm of the coefficients – this term simultaneously reduces the effect of sampling noise and the condition number of the coefficients. We show the effect of this penalization term in numerical simulations: minimax MPF outperforms the well-conditioned MPF and the best product formula.

Finally, we provide a rigorous upper bound on the estimation error of minimax MPF and demonstrate that it scales proportionally to the
aforementioned upper-bound on MPF error. Last but not the least, minimax MPF requires solving small-scale convex optimization problems which can be solved efficiently.   


The rest of this paper is organized as follows.
We provide a necessary background on
product formulas and the theory of Trotter error with commutator scaling
in Section~\ref{sec:PF}, which is largely based on~\cite{childs2021theory}.
Multi-product formulas and our bound on the Trotter errror (Theorem~\ref{thm:MPF}) are stated in Section~\ref{sec:MPF}.
We introduce minimax milti-product formulas
in Section~\ref{sec:DMPF} .
The rigorous upper bound on the Trotter error is compared with
numerical simulations in Section~\ref{sec:numerics}.
This section also reports numerical experiments with static, dynamic and minimax MPFs.
Appendix~\ref{app:A} contains the proofs.

\section{Product  formulas and Trotter error}
\label{sec:PF}

Suppose $H$ is a Hamiltonian describing a quantum spin or fermionic system.
Our goal is to simulate Hamiltonian dynamics governed by the
the von Neumann equation
\be
\label{vonNeumann}
\dot{\rho}(t) = -i[H,\rho(t)], \qquad t\ge 0
\ee
with a  fixed initial state $\rho(0)=\rho_{in}$.
Its solution is
\be
\label{rho(t)}
\rho(t)=e^{-itH} \rho_{in} e^{itH}.
\ee
A common strategy for simulating Hamiltonian dynamics on a quantum computer
relies on product formulas.
Suppose we are given a decomposition
\be
\label{Hdef}
H=\sum_{a=1}^d F_a,
\ee
where $F_a$ are  hermitian operators such that each unitary $e^{-itF_a}$
admits an efficient  implementation by a quantum circuit for any evolution time $t$.
For example, this is the case if $F_a$ is a sum of few-particle interactions
that pairwise commute.
A product formula associated with $F_1,\ldots,F_d$
is  an operator-valued function
\be
\label{PFdef}
\calS(t) = e^{-itF_d} \cdots e^{-itF_2} e^{-itF_1}.
\ee
Condition Eq.~(\ref{Hdef}) ensures that $\calS(t) = e^{-itH} + O(t^2)$ in the limit $t\to 0$.
 More generally,
$\calS(t)$ is called an order-$p$ product formula if
\be
\label{PForder}
\calS(t) =e^{-itH} + O(t^{p+1})
\ee
in the limit $t\to 0$.
For example, suppose $H$ describes a chain of $n$ qubits with
nearest-neighbor interactions, $H=\sum_{j=1}^{n-1} H_{j,j+1}$.
Then a second-order Trotter-type product formula can be chosen as
\be
\label{2nd_order}
\calS(t)= e^{-itF_3} e^{-itF_2} e^{-itF_1}
\ee
with $F_1=F_3 = (1/2)(H_{1,2}+H_{3,4}+H_{5,6}+\ldots)$
and
$F_2=H_{2,3}+H_{4,5}+H_{6,7}+\ldots$.
In this example $d=3$.

Given an integer  $k\ge 1$ and a product formula $\calS(t)$,
define a state $\rho_k(t)$ obtained from $\rho_{in}$ by applying $k$ steps
of the product formula $\calS(t/k)$, that is,
\be
\label{rho_k(t)}
\rho_k(t) = \calS(t/k)^k \rho_{in} \calS(t/k)^{-k}.
\ee
We shall refer to the approximation error $\|\rho_k(t)-\rho(t)\|_1$
achieved by a product formula as a Trotter error.
Recall that the 1-norm $\|X\|_1$
of an operator $X$ is defined as the sum
of singular values of $X$ or, equivalently, as
$\|X\|_1=\max_\calO |\mathrm{Tr}(\calO X)|/\|\calO\|$, where the maximum is
over all non-zero operators $\calO$.
Using the standard properties of the 1-norm~\cite{nielsen2002quantum} and
the triangle inequality one gets
\begin{align}
\label{1norm2operator}
\|\rho_k(t)-\rho(t)\|_1 & \le 2\| \calS(t/k)^k - e^{-itH}\| \nonumber \\
& \le 2k \| \calS(t/k) - e^{-i(t/k)H}\|.
\end{align}
A general upper bound on the error $\| \calS(t)-e^{-itH}\|$
which is nearly tight in many cases of interest
has been recently proved by Childs, Su, Tran, et al.~\cite{childs2021theory}.
To state this bound we need some more notations.
Suppose $A_1,\ldots,A_s,B$ are linear operators acting on the same space.
Define a quantity
\begin{align}
\label{alpha_comm}
\alpha_{\comm}(p;A_1,\ldots,A_s;B)=
\sum_{
\substack{q_1,\ldots,q_s\ge 0 \\q_1+\ldots+q_s=p\\}}
\frac{p!}{q_1!\cdots q_s !} \nonumber\\
\cdot \| (\ad_{A_1})^{q_1} \cdots (\ad_{A_s})^{q_s} (B)\|.
\end{align}
Here $\ad_X$ denotes the adjoint action of an operator $X$, that is, $\ad_X(Y)=XY-YX$
for all operators $Y$.
Assuming that the operators $A$'s and $B$ have the units of energy,
 $\alpha_{\comm}(p;\ldots)$ has units
$[\mathrm{energy}]^{p+1}$.
The following lemma is a corollary of the upper bound established in~\cite{childs2021theory}.
\begin{lemma}[\bf Trotter error]
\label{fact:trotter_error}
Let $\rho(t)$ and $\rho_k(t)$ be the exact time evolved
state and its approximation obtained by applying $k$ steps
of an order-$p$ product formula, see Eqs.~(\ref{rho(t)},\ref{rho_k(t)}).
Then for all $t\ge 0$
\be
\label{childs_su}
\| \rho_k(t) - \rho(t) \|_1 \le \frac{2\alpha_p t^{p+1}}{(p+1)! k^p}
\ee
where
\be
\label{alpha_p}
\alpha_p=\sum_{a=2}^d \alpha_{\comm}(p;F_d,\ldots,F_a; F_{a-1}).
\ee
\end{lemma}
Since our setting is different from the one of
Ref.~\cite{childs2021theory},
we provide a proof of Lemma~\ref{fact:trotter_error}  in Appendix~\ref{app:A}.
As a concrete example, consider the
second-order Trotter-type product formula Eq.~(\ref{2nd_order}).
Then Eq.~(\ref{alpha_p}) gives
\[
\alpha_2=\| \, [F_2,[F_2,F_1]]\, \| + 3 \| \, [F_1,[F_1,F_2]]\, \|.
\]
In general, $\alpha_p$ involves the norm of order-$p$ nested commutators
composed of $F_1,\ldots,F_d$.
The quantity $\alpha_p$ scales linearly with the system size $n$ for many interesting Hamiltonians
such as quantum lattice models with short-range interactions~\cite{childs2021theory}.
In this case  the Trotter error Eq.~(\ref{childs_su}) is proportional to $n t^{p+1}/k^p$ with a constant
factor that depends on the order $p$ and details of the considered Hamiltonian.

\section{Multi Product Formulas}
\label{sec:MPF}

A multi product formula (MPF)
approximates the exact time evolved state $\rho(t)$ by a linear combination
\be
\label{MPF}
\mu(t) = \sum_{i=1}^r c_i \rho_{k_i}(t)
\ee
where $\rho_{k_i}(t)$ is an approximation of $\rho(t)$
obtained by applying $k_i$ steps of some base product formula $\calS(t)$,
see Eq.~(\ref{rho_k(t)}), and  $c_i$ are real coefficients.
All terms in $\mu(t)$  use the same base product formula.
The key idea behind MPFs is that  errors
introduced by each  individual term in $\mu(t)$ can be
approximately cancelled with a proper choice of the coefficients
$c_i$. Thus the error $\|\mu(t)-\rho(t)\|_1$ achieved by an MPF
can be much smaller than the errors $\|\rho_{k_i}(t)-\rho(t)\|_1$
achieved by each term.
We are interested in the error $\|\mu(t)-\rho(t)\|_1$
minimized over the coefficients $c_i$. Note that the optimal
MPF $\mu(t)$ might not be a physical state since some coefficients
$c_i$ can be negative.
 Accordingly, one may not be able to prepare
$\mu(t)$ in the lab. However, this is not needed if one's goal is just
to compute an expected value of some observable $\calO$ on $\rho(t)$.
Indeed, suppose one can efficiently prepare each individual state
$\rho_{k_i}(t)$ and obtain an estimate $x_i\in \RR$ satisfying
\be
\label{mean_value_error1}
\left| \mathrm{Tr}(\calO \rho_{k_i}(t)) -x_i\right|\le \epsilon_i
\ee
for some error tolerance $\epsilon_i$.
 Assuming $\|\calO\|\le 1$ one gets
\begin{align}
\left| \mathrm{Tr}(\calO \rho(t)) - \sum_{i=1}^r c_i x_i \right| & \le
\|\mu(t)-\rho(t)\|_1 \nonumber \\
& + \sum_{i=1}^r \epsilon_i |c_i|.
\label{mean_value_error2}
\end{align}
The last term in Eq.~(\ref{mean_value_error2}) can be made arbitrarily small
 simply by
improving the quality of estimates in Eq.~(\ref{mean_value_error1}).
However, to avoid error amplification, the MPF must be well-conditioned~\cite{low2019well}
such that the condition number
$\kappa= \sum_{i=1}^r |c_i|$
 is sufficiently small.

Let us first consider the
error $\|\mu(t)-\rho(t)\|_1$. We stress that the existing rigorous error bounds for MPFs
may not  be applicable in the regime covered by
Lemma~\ref{fact:trotter_error}.
Indeed, assuming that the bound of Lemma~\ref{fact:trotter_error}
scales as  $O(nt^{p+1}/k^p)$, where $n$ is the number of qubits,
the base product formula needs only $k=\Omega(n^{1/p} t^{1+1/p})$
time steps for an accurate simulation.
Meanwhile, the existing error bounds for MPFs such as Lemma~10
of Ref.~\cite{rendon2022improved} are only applicable if $\min_j k_j=\Omega(nt)$,
assuming that the Hamiltonian $H$ contains $\Omega(n)$
terms with the norm $\Omega(1)$.
Applying such bounds to quantum advantage demonstrations
in which $t$ is comparable or smaller than $n$
for a suitable choice of energy units~\cite{childs2018toward,kim2023scalable}
would require
each term in the MPF to perform at least  $\Omega(n^2)$ time steps,
whereas $\Omega(n^{1+2/p})$ steps would suffice for the base
product formula. Note that $n^{1+2/p}\ll n^2$ for $p\ge 3$.

To justify the use of MPFs in the regime covered by Lemma~\ref{fact:trotter_error},
it is desirable
to extend the theory of Trotter error with commutator scaling developed
in~\cite{childs2021theory} from regular product formulas to MPFs.
This is achieved in the following theorem.
Our bound on the Trotter error depends on the norm of nested commutators
analogous to $\alpha_{\comm}$
and $\alpha_p$, see Eqs.~(\ref{alpha_comm},\ref{alpha_p}). Let us first define these commutators.
Fix a product formula
$\calS(t)=e^{-itF_d}\cdots e^{-it_1 F_1}$
and consider a set of unitary operators
\[
\Gamma(t) = \{ e^{-i\tau_d F_d} \cdots e^{-i\tau_1 F_1}\, \vert\,
0\le \tau_1,\ldots,\tau_d\le t\}.
\]
Here we assume $t\ge 0$.
Given a unitary $U\in \Gamma(t)$,
let $\hat{U}$ be a linear map such that $\hat{U}(X)=UXU^{-1}$ for any operator $X$.
Define a quantity
\begin{align}
\label{beta_comm1}
\beta_{\comm}(p,\ell;A_1,\ldots,A_s;B;t)=
\sum_{
\substack{q_1,\ldots,q_s\ge 0 \\q_1+\ldots+q_s=p\\}}
\frac{p!}{q_1!\cdots q_s !} \nonumber\\
\max_{U\in \Gamma(t)}
\cdot \| (\ad_{H})^\ell \hat{U} (\ad_{A_1})^{q_1} \cdots (\ad_{A_s})^{q_s} (B)\|.
\end{align}
Assuming that all operators $A$'s and $B$ have the units of energy,
$\beta_{\comm}(p,\ell;\ldots)$ has units
$[\mathrm{energy}]^{p+\ell+1}$.
Let
\be
\label{beta_comm2}
\beta_{p,\ell}(t) = \sum_{a=2}^d \beta_{\comm}(p,\ell; F_d,\ldots,F_a; F_{a-1};t).
\ee
Our main result is as follows.
\begin{theorem}[\bf MPF Trotter error]
\label{thm:MPF}
Let $\rho(t)$ and $\rho_{k}(t)$ be the exact time evolved state
and its approximation obtained by applying $k$ steps of an order-$p$ product formula.
Consider
a multi product formula $\mu(t)=\sum_{i=1}^{r} c_i \rho_{k_i}(t)$
with $r=p+1$.
Suppose  the coefficients $c_i$ solve a linear system
\be
\label{ZNE}
\sum_{i=1}^r c_i =1
\quad \mbox{and} \quad
\sum_{i=1}^r \frac{c_i}{k_i^q} = 0
\ee
for $q\in \{p,p+1,\ldots,2p-1\}$.
Then for all $t\ge 0$
\[
\| \mu(t) - \rho(t)\|_1 \le \left( \sum_{i=1}^r \frac{|c_i|}{k_i^{2p}}\right)
 ( a_1 t^{2p+2} + a_2 t^{2p+1} + a_3 t^{2p} ),
\]
where
\[
a_1 =8 \left(\frac{\alpha_p}{(p+1)!}\right)^2,
\]
\[
a_2 = \frac{4\beta_{2p,0}(0)}{(2p)!}
+\frac{8 \beta_{p,p}(t/k_{min})}{(2\pi)^p p!},
\]
and
\[
a_3 =4\sum_{\ell=1}^p \frac{B_\ell \beta_{2p-\ell,\ell-1}(t/k_{min})}{\ell! (2p-\ell)!}.
\]
Here
$B_\ell$ is the $\ell$-th Bernoulli number and
$k_{min}=\min{(k_1,\ldots,k_r)}$.
\end{theorem}

\begin{figure*}
 \centering
 \includegraphics[width=12cm]{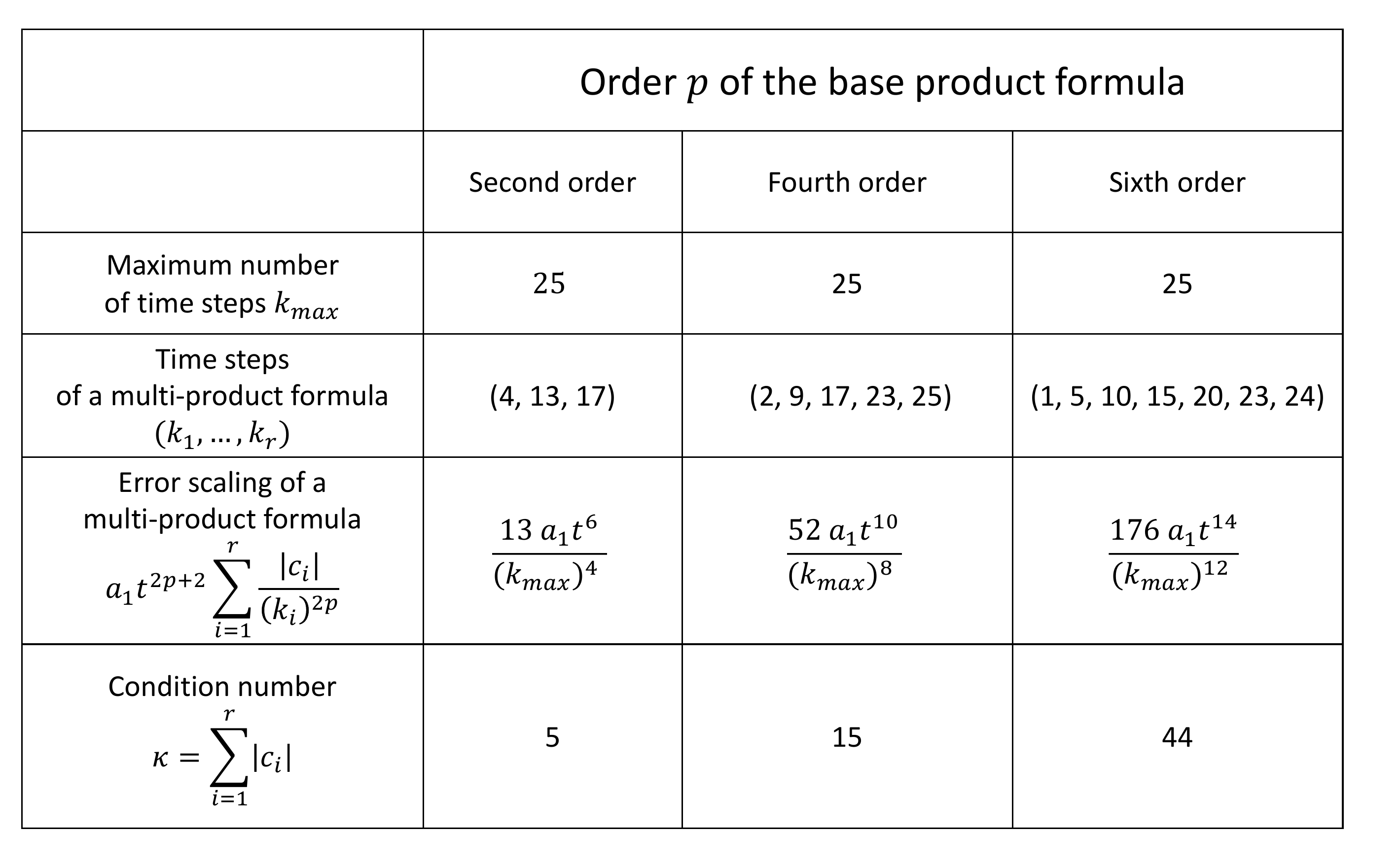}
\caption{Examples of multi-product formulas (MPFs)
of order $p=2,4,6$. Such MPFs have a form $\mu(t)=\sum_{i=1}^{p+1} c_i \rho_{k_i}(t)$
where $\rho_{k_i}(t)$ is an approximation of the exact time evolved state
obtained by applying $k_i$ steps of the base product formula $\calS(t/k_i)$.
The sequence of time steps $(k_1,\ldots,k_{p+1})$ reported in the table was obtained
by minimizing the factor $\sum_{i=1}^{p+1} |c_i|/k_i^{2p}$ in the upper bound of Theorem~\ref{thm:MPF}
over all tuples $(k_1,\ldots,k_{p+1})$ with $1\le k_i\le k_{max}$
and solving the linear system Eq.~(\ref{ZNE}) with $r=p+1$ to obtain the coefficients $c_i$.
To avoid clutter, we round the condition number and coefficients in the error scaling
to the nearest integer. Note that the MPF approximation error can be reduced by performing
a rescaling $k_i\gets \lambda k_i$ for all $i$, where
$\lambda\ge 1$ is any integer. The reduces the
upper bound of Theorem~\ref{thm:MPF}
 by the factor $1/\lambda^{2p}$
without changing the coefficients $c_i$ and the condition number.
Meanwhile, for the regular order-$p$ product formula, the rescaling $k\gets \lambda k$
reduces the approximation error only by the factor $1/\lambda^p$,
see Lemma~\ref{fact:trotter_error}.
}
\label{fig:table_examples}
\end{figure*}

In the limit of large evolution time $t$
the upper bound of Theorem~\ref{thm:MPF} becomes
\be
\label{MPFerror_simple}
\| \mu(t) - \rho(t)\|_1 \le \left( \sum_{i=1}^r \frac{|c_i|}{k_i^{2p}}\right)
a_1 t^{2p+2}\left[ 1+O(1/t)\right].
\ee
Comparing this and Eq.~(\ref{childs_su}) one
infers that the MPF
achieves a nearly quadratic reduction of error compared with the base product formula
for large $t$. To quantify this, we need
to choose an explicit sequence
$(k_1,\ldots,k_r)$  and solve the linear system Eq.~(\ref{ZNE})
to obtain the coefficients $c_i$.
Figure~\ref{fig:table_examples}
shows examples of MPFs  based on product formulas
of order $p=2,4,6$ with $r=p+1$ terms and
the maximum number of time steps per circuit
 $k_{max}=25$.
A sequence of time steps $(k_1,\ldots,k_r)$
was selected by  minimizing the sum  $\sum_{i=1}^r |c_i|/k_i^{2p}$\
over all $r$-tuples $(k_1,\ldots,k_r)$ with $1\le k_i\le k_{max}$.
Each of these examples can be turned to a family of MPFs by a rescaling
$k_i\gets \lambda k_i$, where $\lambda\ge 1$ is any integer.
This reduces the MPF error by the factor $1/\lambda^{2p}$ without changing
the coefficients $c_i$ and
the condition number $\kappa$ (since the latter depend only on the ratios
$k_i/k_j$).
In contrast, applying the rescaling $k\gets \lambda k$ to the base
product formula reduces the approximation error only by the factor $1/\lambda^p$,
see Lemma~\ref{fact:trotter_error}.
Thus  MPFs with a constant condition number $\kappa=O(1)$
achieve a quadratic error reduction compared with the
base product formula without increasing the maximum number
of time steps per circuit (and thus without increasing the circuit depth).
The number of circuit repetitions grows only by a constant factor since
the condition number amplifies the approximation error in Eq.~(\ref{mean_value_error2}).

The scaling of coefficients $a_2$ and $a_3$ in Theorem~\ref{thm:MPF} with the number of qubits $n$ may depend on
details of the Hamiltonian $H$ and the base  product formula.
In Appendix~\ref{app:linear_scaling_with_n} we show that $a_2$ and $a_3$ scale at most linearly with $n$ for a large class of quantum spin Hamiltonians.
This class includes
any quantum lattice model with short-range bounded-norm interactions.
However, we have to assume that the order $p$ and the number of exponents $d$ in the product formula are constants independent of $n$.

We defer the proof of Theorem~\ref{thm:MPF}  to Appendix~\ref{app:A}.
The proof relies  on the machinery introduced in Ref.~\cite{childs2021theory}.
Our main innovation is an integral representation of MPFs
based on
Euler–Maclaurin summation formula. Informally, this integral representation reduces the
problem of bounding the MPF error to a single time step
such that the error operator $\rho_{k_j}(t)-\rho(t)$ can be
viewed as a function of the variable $t/k_j$ only.
We consider Taylor series for this function at the point $t/k_j=0$.
Terms of order less than $2p$ in the Taylor series
give a zero contribution to
 $\mu(t)$
 due to Eq.~(\ref{ZNE}) and the
assumption that the base product formula has order $p$. The norm of the remaining terms of order at least $2p$ can be bounded
by the quantities proportional to $(\alpha_p)^2$ or
$\beta_{p_1,p_2}$
using Theorem~5  of Ref.~\cite{childs2021theory}.
This theorem
bounds the norm of truncated Taylor
series for operator-valued functions such as $e^{-it\ad_{F_d}}\cdots e^{-it\ad_{F_a}}(F_{a-1})$.
Unfortunately, this theorem
does not apply to our problem out of the box because
 Euler–Maclaurin formula generates several unwanted terms proportional to
derivatives of the considered function. To show that these derivative
terms are sufficiently small,
we generalize
Theorem~5  of Ref.~\cite{childs2021theory},
as stated in Lemma~\ref{lemma:remainder} in Appendix~\ref{app:A}.

We expect that the upper bound of Theorem~\ref{thm:MPF} can be strengthened
to give a better than quadratic error suppression
in the case $r>p+1$
and leave this extension for a future work.

\section{Minimax Multi-Product Formulas}
\label{sec:DMPF}
In what follows we refer to MPF formula $\mu^S(t)=\sum_{i=1}^r c_i \rho_{k_i}(t)$ with static $c_i$ and $\rho_{k_i}(t)$ defined as in the previous section as \emph{static MPF}.
By design, $\mu^S(t)$ reduces the Trotter error
compared with the regular product formulas
but this error is not minimized, i.e. $\mu^S(t)$ does not represent the optimal projection of $\rho(t)$ onto MPF subspace, namely the subspace generated by product formulas $\rho_{k_i}(t)$, $i=1\dots r$. Noting that $\|\rho-\mu^S\|_F\le\|\rho-\mu^S\|_1\le\sqrt{r+1}\|\rho-\mu^S\|_F$ in what follows we focus on
minimizing the following projection error:
\be
\| \rho(t)-\mu^D(t)\|_F^2 = \mathrm{Tr}\left( (\rho(t)-\mu^D(t))^2\right)
\ee
over coefficients $c_i$ for each time $t$. In other words, $\mu^D(t) = \sum_{i=1}^r \oc_i(t) \rho_{k_i}(t)$ is the optimal projection of $\rho$ onto MPF subspace in Frobenious norm. In what follows we refer to $\mu^D$ as a \emph{dynamic MPF}. A simple algebra gives
\begin{align}
\label{proj_error}
\| \rho(t)-\mu^D(t)\|_F^2 &=  1 + \sum_{i,j=1}^r M_{i,j}(t) c_i(t) c_j(t) \nonumber \\
&- 2\sum_{i=1}^r L_i^{\exact}(t) c_i(t)
\end{align}
where $M(t)$ is the Gram matrix
\begin{align}\label{eq:Mt}
M_{i,j}(t) & = \mathrm{Tr}(\rho_{k_i}(t) \rho_{k_j}(t)) \nonumber \\
& = \left| \la \psi_{in} |\calS(t/k_i)^{-k_i} \calS(t/k_j)^{k_j}|\psi_{in}\ra\right|^2
\end{align}
and
$L^{\exact}(t)$ is a vector of overlaps
\begin{equation}\label{eq:Lexact}
L_i^{\exact}(t) = \mathrm{Tr}(\rho(t) \rho_{k_i}(t)).
\end{equation}
If we knew both $M(t)$ and $L^{\exact}(t)$ then finding $\oc$ by minimizing the
right-hand side of Eq.~(\ref{proj_error}) over the coefficients $c(t)$ is straightforward. However, even though the Gram matrix $M(t)$ can be efficiently measured on hardware
(with a small additive error), the vector of overlaps $L^{\exact}(t)$ is unknown
since it depends on the the exact solution $\rho(t)$. To overcome this we approximate $L_i^{\exact}$ and focus on estimating $\oc$ by taking the approximation error into account.

\paragraph{Algorithmic error.} Assume that we need to approximate $\rho(t)$ on a segment $[t_0,T]$. Fix time step $dt>0$ and select a uniform grid on $[t_0,T]$: $t_j=t_0+j\,dt$, with integer $j=0,\dots,(T-t_0)/dt$. We approximate $\rho(t_{j+1})$ by
\[
\rho^{\apx}(t_{j+1})=\calS(dt/k_0)^{k_0} \mu^D(t_j)\calS(dt/k_0)^{-k_0}
\] for an integer $k_0>0$. Then $L^{\exact}_i(t_{j+1})$ is approximated by
\be\label{eq:Lapprox}
L_i^{\apx}(t_{j+1})= \tr\left( \rho^{\apx}(t_{j+1}) \rho_{k_i}(t_{j+1})\right).
\ee
and by linearity of the trace,
\be
\label{Lapprox}
L_i^\apx(t_{j+1}) = \sum_{s=1}^r Q_{i,s}(t_{j+1}) \oc_s(t_j),
\ee
where
\begin{align}\label{eq:Qt}
Q_{i,s}&(t_{j+1})  =  \tr\left(\calS(\frac{dt}{k_0})^{k_0} \rho_{k_s}(t_j) \calS(\frac{dt}{k_0})^{-k_0}
 \rho_{k_i}(t_{j+1})\right)\nonumber \\
 &
= \left| \la \psi_{in} |\calS(\frac{t_{j}}{k_s})^{-k_s}  \calS(\frac{dt}{k_0})^{-k_0} \calS(\frac{t_{j+1}}{k_i})^{k_i}|\psi_{in}\ra\right|^2
\end{align}
is a matrix that we can efficiently compute on hardware (with a small additive error). Making use of the above approximation we find that $\oc(t_j)$ solves: $M(t_j)\oc(t_j) = Q(t_j) \oc(t_{j-1}) + L^{\exact}(t_j) -  L^{\apx}(t_{j})$. To enforce $\tr(\rho^{\apx}(t))=1$ we require an extra constraint: $\mathbb{1}^\top \oc(t)=1$ where $\mathbb{1}$ is the $r$-dimensional vector of ones.

\paragraph{Shot noise.} In what follows we assume that the "exact" matrices $M(t)$ and $Q(t)$, defined by~\cref{eq:Mt,eq:Qt} respectively, belong to an ellipsoid centered around the given matrices $\bar M(t)$ and $\bar A(t)$:

\begin{align}
  \label{eq:MQellips}
  M(t) &= \bar M(t) + E(t)^\star_1,\quad  Q(t)= \bar A(t) + E(t)^\star_2
\end{align}
with
\be
\|E(t)^\star_{1,2}\|_{2,2}\le \varepsilon
\ee
%
for some unknown $E^\star_{1,2}(t)$ modelling hardware noise of the given magnitude $\varepsilon$ in terms of $\|E(t)\|_{2,2}=\max_{\|x\|_2=1}\|E(t)x\|_2$. In other words: $M(t)\in\{\bar M(t)+E^D_{1}(t), \|E^D_{1}(t)\|_{2,2}\le \varepsilon\}$ and $Q(t)\in\{\bar A(t) + E^D_2(t), \|E^D_{2}(t)\|_{2,2}\le \varepsilon\}$. This assumption implies that $\oc$ in fact solves:
\begin{align}\label{eq:stateq}
  (\bar M(t_j) &+ E^\star_1(t_j)) \oc(t_j) = (\bar A(t_j)+ E^\star_2(t_j)) \oc(t_{j-1}) \\
  &+ L^{\exact}(t_j) -  L^{\apx}(t_{j}),  \mathbb{1}^\top \oc(t)=1\,.
\end{align}
for unknown bounded matrices $E^\star_{1,2}(t)$ changing over time.

\paragraph{Minimax MPF.} In order to estimate $\oc(t)$ we need means to estimate solution of~\cref{eq:stateq} which accounts for the error of approximating $L_i^{\exact}$ by $L^{\apx}_i$ and the measurement error~\eqref{eq:MQellips}. To this end we introduce the minimax MPF algorithm. The key difference to the dynamic MPF algorithm described above is the inclusion of a penalization term $\varepsilon \|x\|_2$ - see Algorithm \ref{alg:mdMPF} below.
\begin{algorithm}
\caption{Minimax MPF}\label{alg:mdMPF}
\SetKwInOut{Input}{input}
\SetKwInOut{Output}{output}
\Input{window $[t_0,T]$, $\oc(t_0)$, time step $dt>0$, noise magnitude $\varepsilon$, and matrices $\bar M(t_j)$, $\bar A(t_j)$ for $t_j=t_0+j dt$, $j=0,\dots,(T-t_0)/dt$ (see~\cref{eq:MQellips})}
\Output{estimate $\hat c(t_j)$ of $\oc(t_j)$}
\SetKwBlock{Beginn}{beginn}{ende}
\Begin{Set $\hat c(t_0)=\oc(t_0)$\;
  \For{$j=1$ to $(T-t_0)/dt$}{
    $t_j:= t_0+j dt$, $\hat c(t_j):=\hat x$;\
   $\hat x\in \operatorname{Argmin}_{\mathbb{1}^\top x=1} \{\|\bar M(t_j) x - \bar A(t_j)\hat c(t_{j-1})\|_2+\varepsilon \|x\|_2 \}$
        }
} %
\end{algorithm}



\paragraph{Error bounds for minimax MPF.} \label{par:err}
In Appendix \ref{sec:minimax_dae}, we provide a theorem - see \Cref{t:mdMPF-error} -  that quantifies the error
of minimax MPF (Algorithm~\ref{alg:mdMPF}), namely the worst-case error between $\oc(t_j)$ and $\hat c(t_j)$ for time instant $t_j$. \Cref{t:mdMPF-error} demonstrates that the worst-case estimation error of minimax MPF scales proportionally to the upper-bound for static MPF error given in~\ref{thm:MPF}. As one should expect, this theorem shows that when the sampling noise is negligible (i.e. $\varepsilon\approx0$ ) then minimax MPF coincides with dynamic MPF.

\section{Numerical experiments}
\label{sec:numerics}

\begin{figure}[h]
     \centering
     \includegraphics[width=9cm]{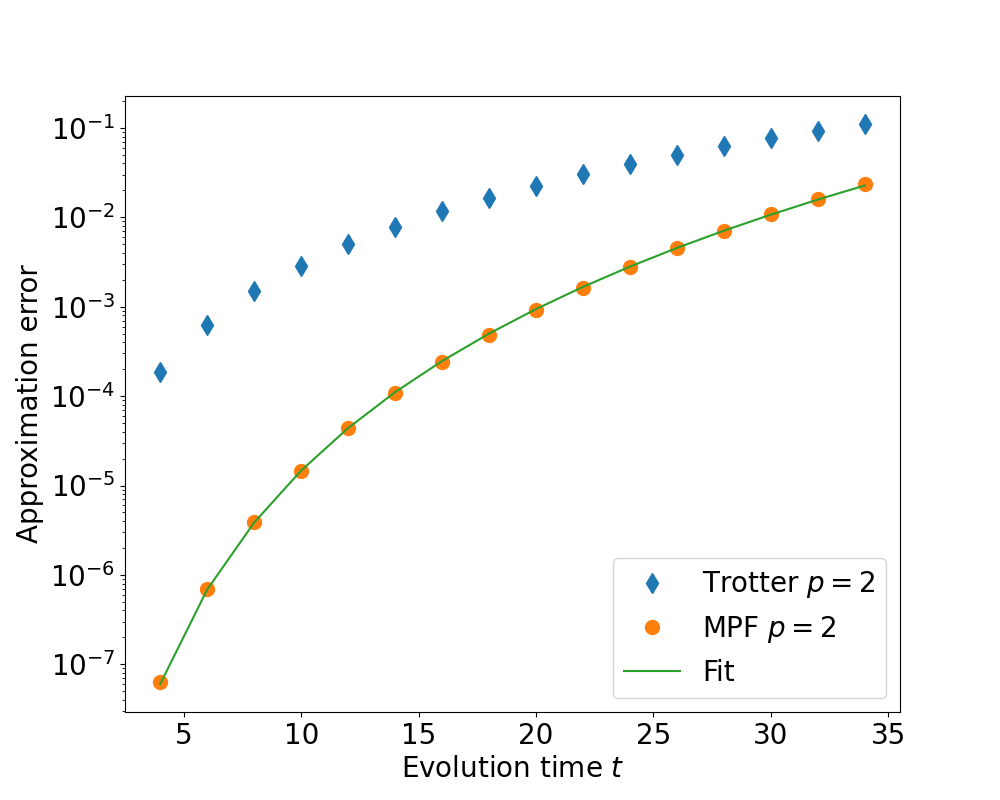}
     \caption{Approximation error achieved by the second-order Trotter circuit with $k_3=850$ time steps (blue)
     and MPF with $(k_1,k_2,k_3)=(200,650,850)$ (orange) for the Heisenberg spin chain Hamiltonian Eq.~(\ref{spin_chain})
     with $n=14$ qubits. Green line shows the fitting
     formula Eq.~(\ref{order2fit}).
     \label{plot:order2}}
\end{figure}

\begin{figure}[h]
     \centering
     \includegraphics[width=9cm]{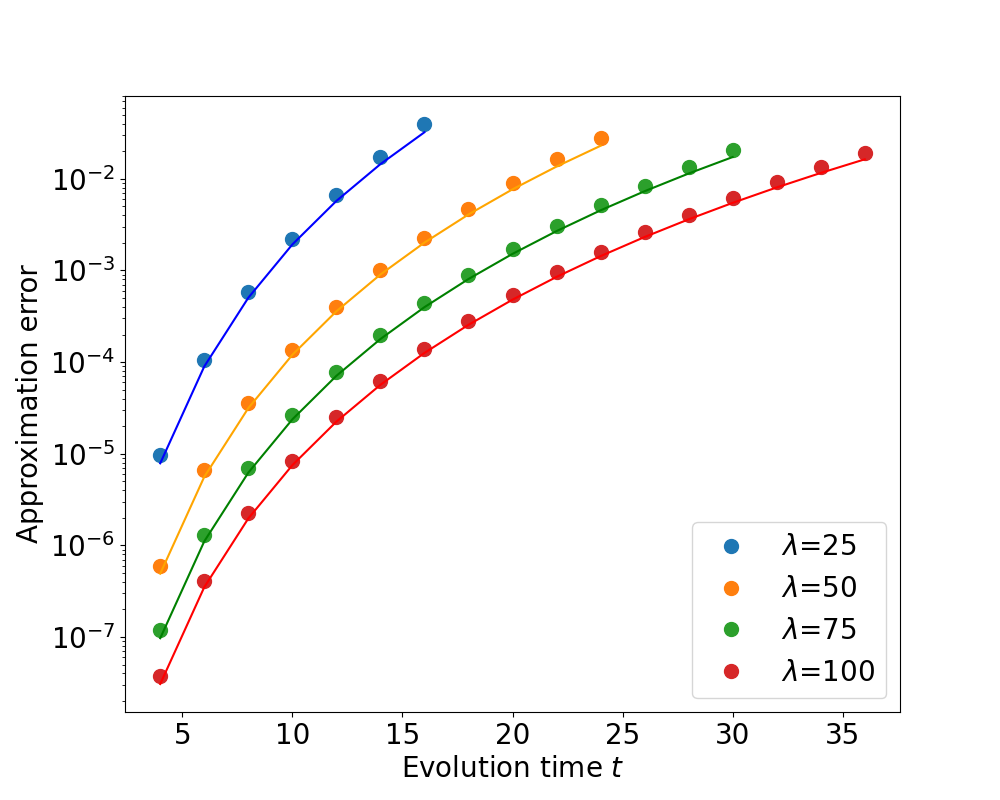}
      \includegraphics[width=9cm]{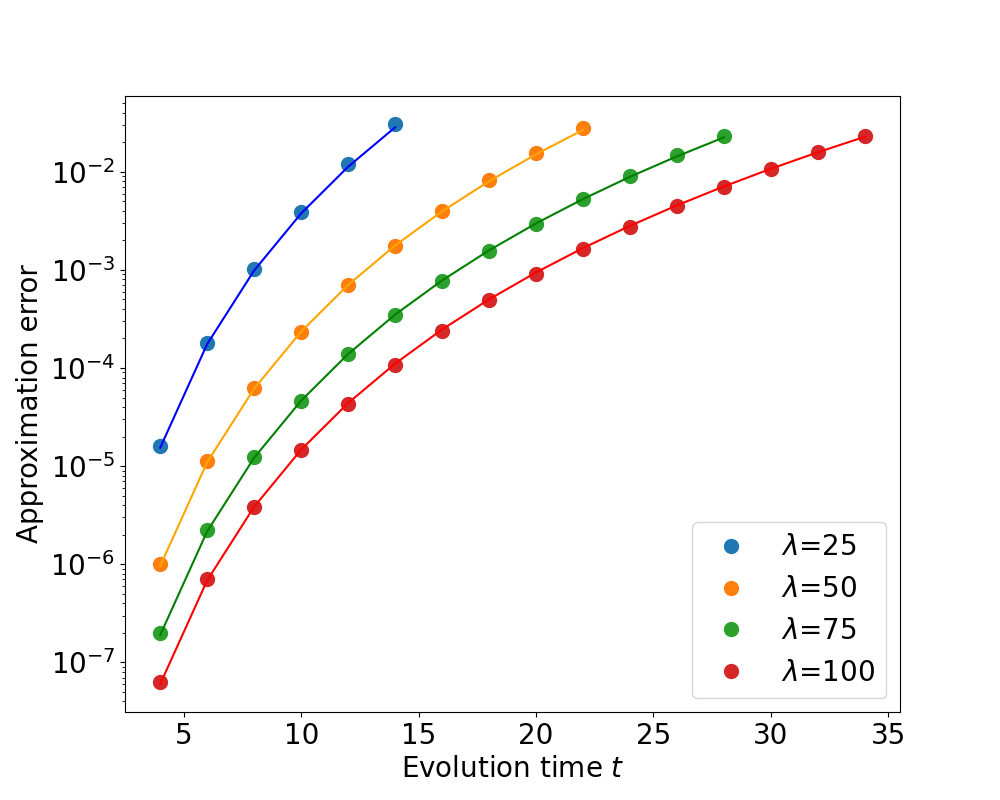}
     \caption{Comparison between the true MPF approximation error $\|\mu(t)-\rho(t)\|_1$ (circles) and fitting formula Eq.~(\ref{order2fit}) (solid  lines) for the second-order product formulas with
     $(k_1,k_2,k_3)=\lambda \cdot (4,13,17)$. Top panel: $n=10$, Bottom panel: $n=14$.}
     \label{plot:order2combined}
\end{figure}

We begin by testing predictions of Theorem~\ref{thm:MPF} numerically
for the second and fourth order Trotter-type product formulas.
We select $H$ to be the spin chain Hamiltonian
proposed by Childs, Maslov et al.~\cite{childs2018toward},
\be
\label{spin_chain}
H = \sum_{j=0}^{n-2} (X_j X_{j+1} + Y_j Y_{j+1} + Z_j Z_{j+1})
     + \sum_{j=0}^{n-1} h_jZ_j.
 \ee
Here $X_j,Y_j,Z_j$ are Pauli operators acting on the $j$-th qubit and
 $h_j\in[-1,1]$ are
coefficients drawn randomly from the uniform distribution.
We choose a second order ($p=2$)
product formula as
\[
\calS_2(t)=e^{-itF_5}e^{-itF_4}e^{-itF_3} e^{-itF_2} e^{-itF_1},
\]
where
\[
F_1=F_5 =\frac12 \sum_{\mathrm{odd}\; j}  X_j X_{j+1} + Y_j Y_{j+1} + Z_j Z_{j+1},
\]
\[
F_2=F_4 = \frac12 \sum_{j=0}^{n-1} h_j Z_j,
\]
and
\[
F_3= \sum_{\mathrm{even}\; j}  X_j X_{j+1} + Y_j Y_{j+1} + Z_j Z_{j+1}.
\]
Let $\rho(t)=e^{-itH}|\psi_{in}\ra\la \psi_{in}|e^{itH}$,
where $|\psi_{in}\ra = |1010\ldots 10\ra$
is Neel-type initial state.
Specializing Theorem~\ref{thm:MPF} and Figure~\ref{fig:table_examples} to $p=2$ gives an MPF
\[
\mu(t)=\sum_{i=1}^3 c_i
\rho_{k_i}(t),
\]
where $\rho_k(t)=\calS_2(t/k)^k|\psi_{in}\ra\la \psi_{in}|
\calS_2(t/k)^{-k}$ and
\[
(k_1,k_2,k_3)=\lambda \cdot(4,13,17)
\]
for an integer $\lambda\ge 1$.
Solving the linear system Eq.~(\ref{ZNE}) gives
coefficients
\[
 c_1=0.016088(4),\;
 c_2=-1.794934(6),\;
 c_3=2.778846(1).
\]
Figure~\ref{plot:order2} shows a comparison between three types of approximation
errors: (1) Trotter error $\|\rho(t)-\rho_3(t)\|_1$ achieved by the best Trotter circuit $\calS_2(t/k_3)^{k_3}$,
(2) MPF error $\|\rho(t)-\mu(t)\|_1$, and (3) fitting ansatz based
on Eq.~(\ref{MPFerror_simple}).
The latter was chosen as
\be
\label{order2fit}
\epsilon_{\mathsf{fit}}= 0.06n^2t^{6} \sum_{i=1}^3 \frac{|c_i|}{k_i^{4}}
\approx
\frac{(7.5 \times 10^{-6} )n^2 t^6}{\lambda^4}.
\ee
Here we fitted the
coefficient $a_1$  in Eq.~(\ref{MPFerror_simple})
by a quadratic function of $n$ obtaining an estimate
$a_1\approx0.06n^2$.
Figure~\ref{plot:order2combined}
demonstrates that the fitting formula Eq.~(\ref{order2fit})
closely approximates
the true MPF error
$\|\rho(t)-\mu(t)\|_1$
for several values of $n$ and $\lambda$.
The analogous fitting ansatz for the Trotter error $\|\rho(t)-\rho_k(t)\|_1$
with $k$  steps of the product formula $\calS_2(t/k)$ was found to be
\be
\label{order2fit_Tro}
\epsilon'_{\mathsf{fit}} = \frac{0.6 n t^3}{k^2}.
\ee

One can use Eq.~(\ref{order2fit}) to estimate the number of Trotter steps (and thus the circuit depth)
for a given simulation task. For example,
solving the benchmark problem of Ref.~\cite{childs2018toward} with $n=t=100$  and error tolerance $\epsilon_{\mathsf{fit}}=10^{-3}$ using the second-order MPF would require
$\lambda\approx 3000$
which translates to $k_{max}=17\lambda \approx 40,000$ Trotter steps.

\begin{figure}[h]
     \centering
     \includegraphics[width=9cm]{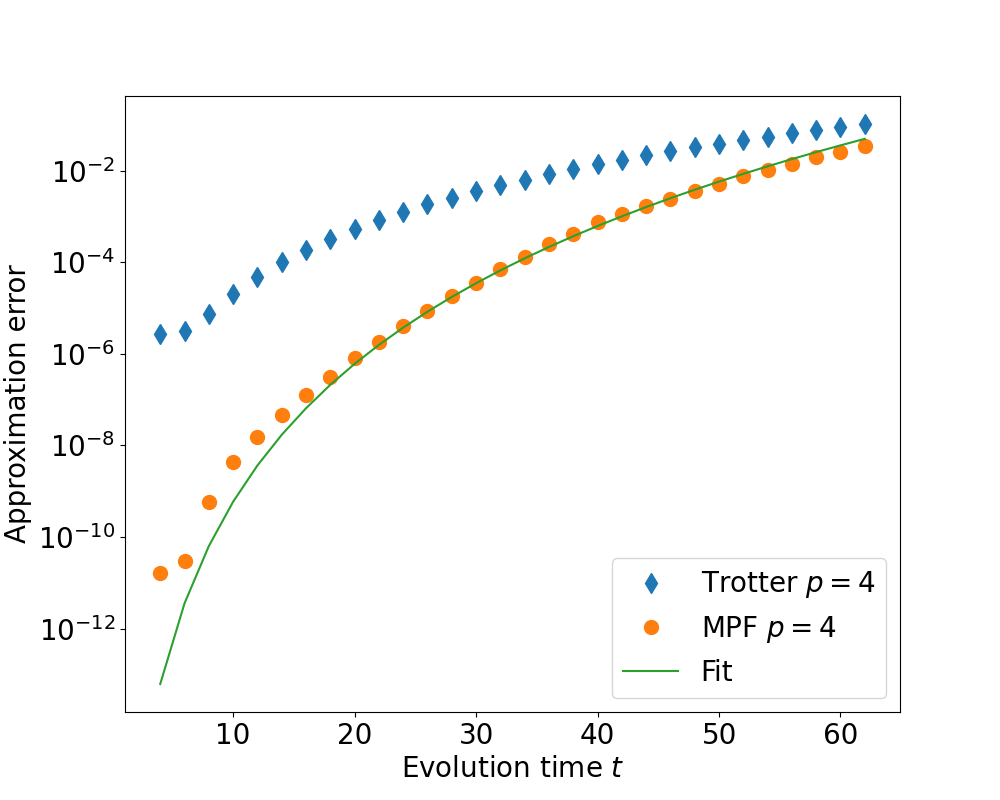}
     \caption{Approximation error achieved by the fourth-order Trotter circuit with $k_5=250$ time steps (blue)
     and MPF with $(k_1,k_2,k_3,k_4,k_5)=(20, 90, 170, 230, 250)$ (orange) for the Heisenberg spin chain Hamiltonian Eq.~(\ref{spin_chain})
     with $n=14$ qubits. Green line shows the fitting
     formula Eq.~(\ref{order4fit}).
     \label{plot:order4}}
\end{figure}

\begin{figure}[h]
     \centering
 \includegraphics[width=9cm]{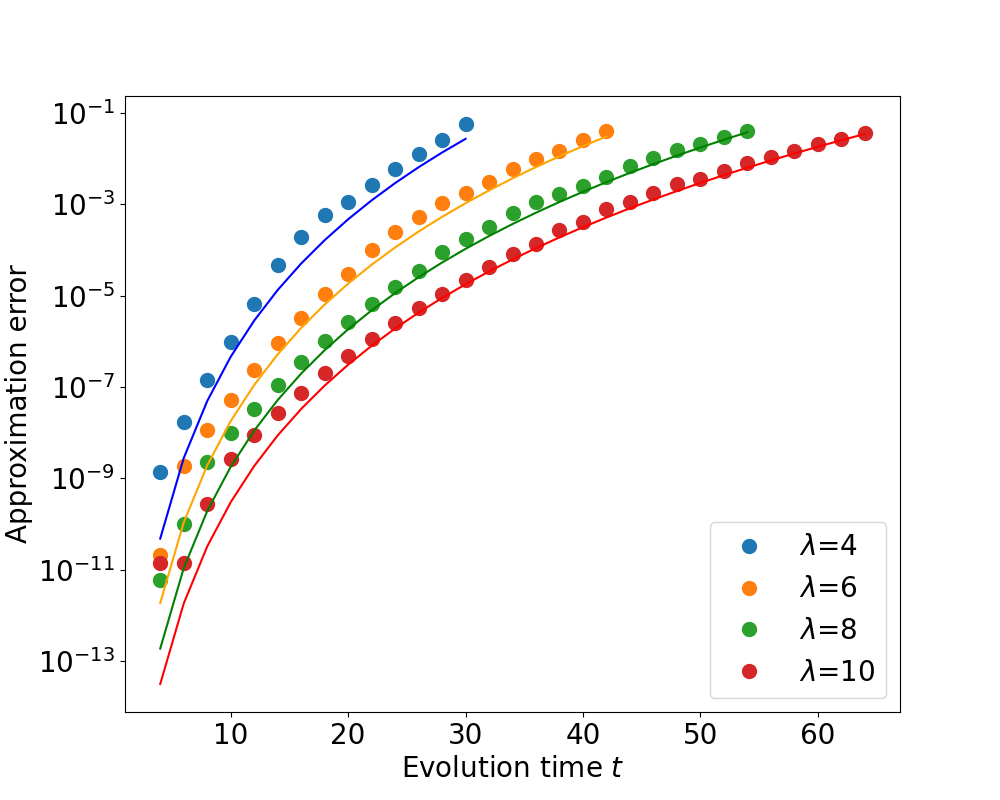}
    \includegraphics[width=9cm]{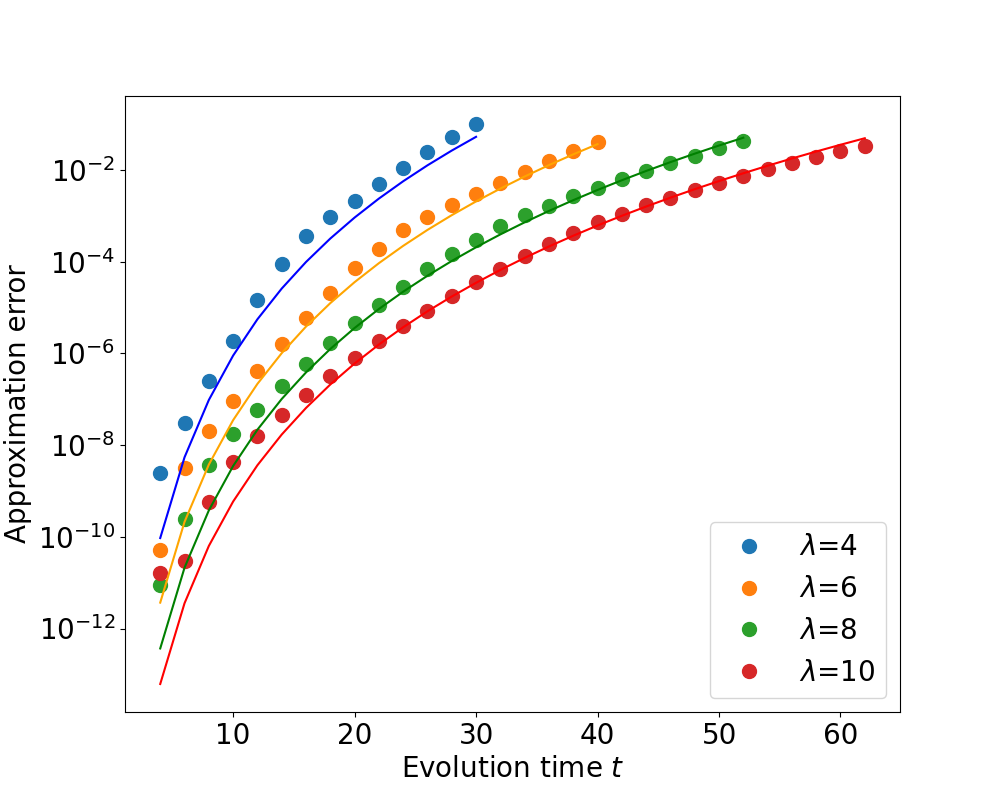}
     \caption{Comparison between the true MPF approximation error $\|\mu(t)-\rho(t)\|_1$ (circles) and fitting formula Eq.~(\ref{order4fit}) (solid  lines) for the fourth-order product formulas with
$(k_1,k_2,k_3,k_4,k_5)=\lambda(2, 9, 17, 23, 25)$. Top panel: $n=10$, Bottom panel: $n=14$.
   For small evolution time $t$ the fitting formula underestimates the error since it neglects
   corrections $a_2t^{2p+1}$ and $a_3t^{2p}$ in the upper bound of Theorem~\ref{thm:MPF}.
   }
\label{plot:order4combined}
\end{figure}

Next, consider the fourth-order ($p=4$) product formula
\[
\calS_4(t) = (\calS_2(u t))^2 \calS_2((1-4u)t)  (\calS_2(u t))^2,
\]
where $u=\frac1{4-4^{1/3}}$,
see for instance~\cite{childs2018toward}. Specializing Theorem~\ref{thm:MPF} and Figure~\ref{fig:table_examples} to $p=4$ gives an MPF
\[
\mu(t)=\sum_{i=1}^5 c_i
\rho_{k_i}(t),
\]
where
\[
(k_1,k_2,k_3,k_4,k_5)=\lambda\cdot (2, 9, 17, 23, 25)
\]
for an integer $\lambda\ge 1$.
Solving the linear system Eq.~(\ref{ZNE}) gives
coefficients
\[
\ba{cc}
 c_1=1.77273114\times 10^{-8} & c_4 = -6.7785310(5)\\
 c_2= -0.00267812(5)  & c_5  =  7.2808414(4)\\
 c_3= 0.50036771(6)  &   \\
 \ea
\]
Figure~\ref{plot:order4} shows a comparison between three types of approximation
errors: (1) Trotter error $\|\rho(t)-\rho_5(t)\|_1$ achieved by the best Trotter circuit $\calS_4(t/k_5)^{k_5}$,
(2) MPF error $\|\rho(t)-\mu(t)\|_1$, and (3) fitting ansatz based
on Eq.~(\ref{MPFerror_simple}).
The latter was chosen as
\be
\label{order4fit}
\epsilon_{\mathsf{fit}}=  0.00014n^2t^{10}  \sum_{i=1}^5 \frac{|c_i|}{k_i^{8}}
\approx \frac{(3\times 10^{-14}) n^2 t^{10}}{\lambda^8}.
\ee
Here we fitted the
coefficient $a_1$  in Eq.~(\ref{MPFerror_simple})
 by a quadratic function of $n$ obtaining an estimate
$a_1\approx 0.00014n^2$.
Figure~\ref{plot:order4combined}
demonstrates that the fitting formula Eq.~(\ref{order4fit})
closely approximates
the true MPF error
$\|\rho(t)-\mu(t)\|_1$
for several values of $n$ and $\lambda$ and large evolution time $t$.
Meanwhile, the fitting formula underestimates the error for small $t$ since it neglects
  the corrections  $a_2t^{2p+1}=a_2t^9$ and $a_3t^{2p}=a_3t^8$ in the upper bound of Theorem~\ref{thm:MPF}. We expect that these correction may become
 more important for higher-order formulas.

The analogous fitting ansatz for the Trotter error $\|\rho(t)-\rho_k(t)\|_1$
with $k$  steps of the product formula $\calS_4(t/k)$ was found to be
\be
\label{order4fit_Tro}
\epsilon'_{\mathsf{fit}} = \frac{0.04n t^5}{k^4}.
\ee

One can use Eq.~(\ref{order4fit}) to estimate the number of Trotter steps
for a given simulation task. For example,
solving the benchmark problem of Ref.~\cite{childs2018toward} with $n=t=100$  and error tolerance $\epsilon_{\mathsf{fit}}=10^{-3}$ using the fourth-order MPF would require
$\lambda\approx 50$
which translates to $k_{max}=25\lambda \approx 1,250$ Trotter steps,
as opposed to $40,000$ Trotter steps for the second-order MPF.
Note however that each Trotter step of $\calS_4$ has the same depth as five
Trotter steps of $\calS_2$. Thus the fourth-order MPF achieves roughly
six-fold depth reduction compared with the second-order MPF.

\begin{figure}[h]
     \centering
     \includegraphics[width=9cm]{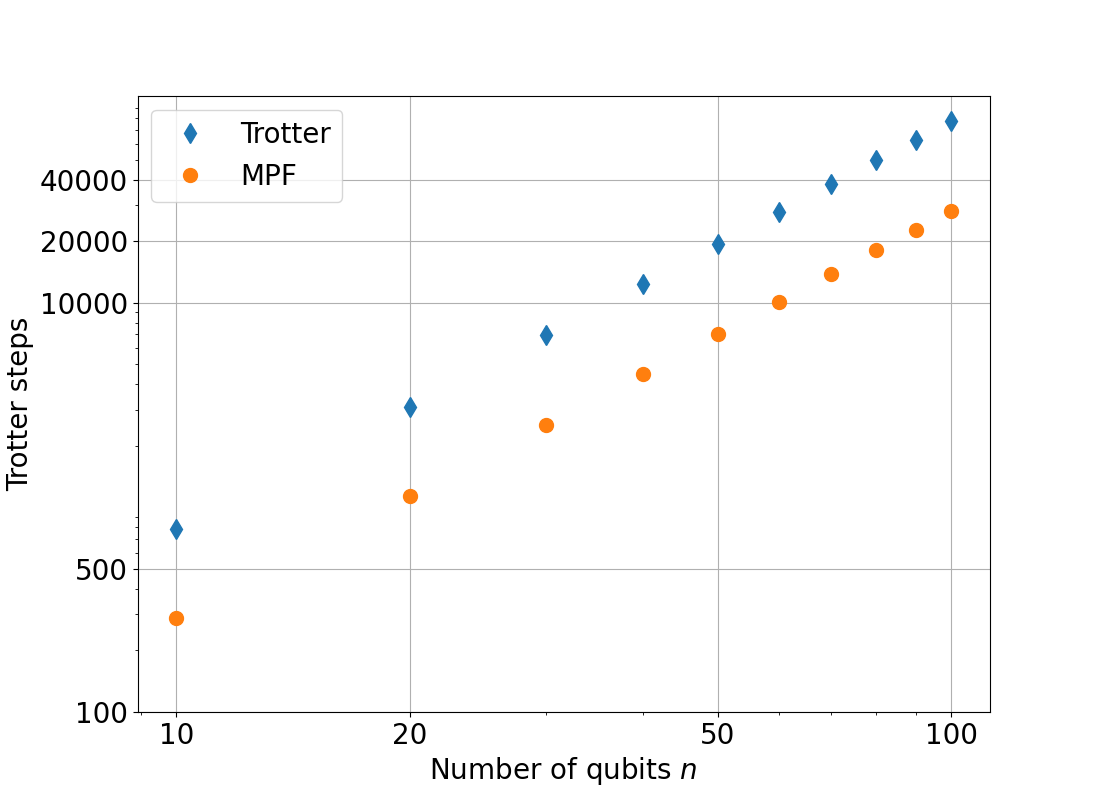}
      \includegraphics[width=9cm]{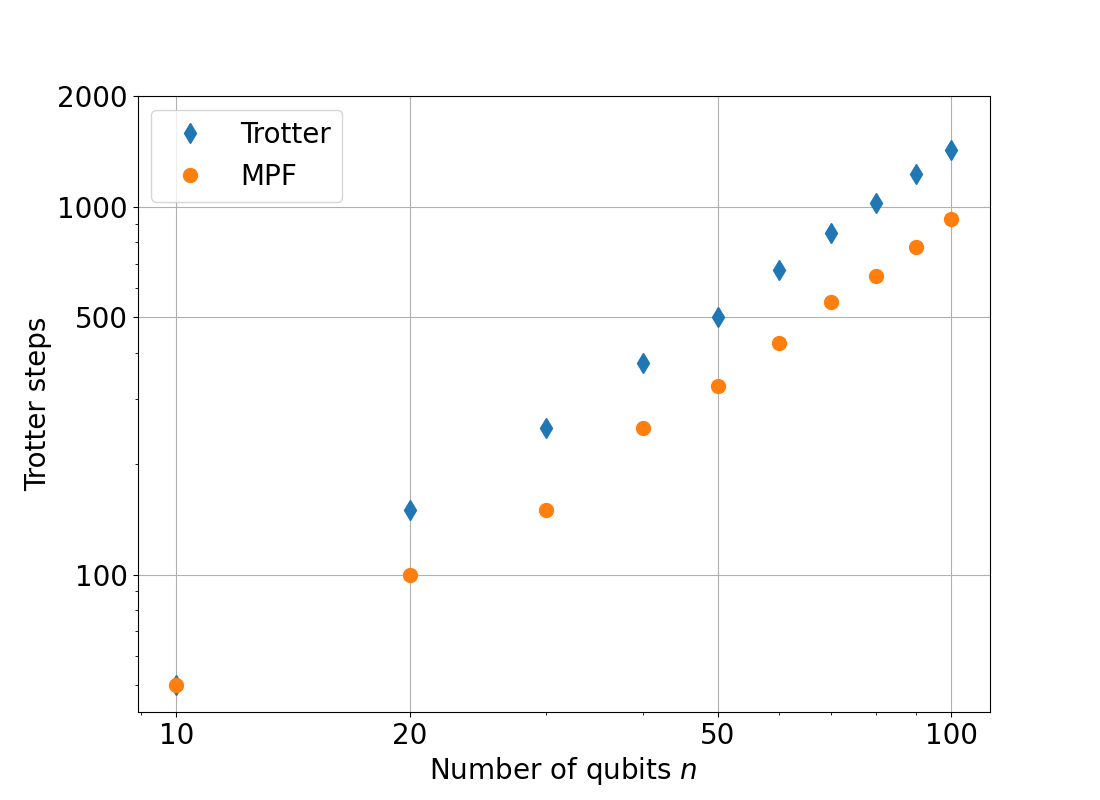}
\caption{The cost of $t=n$ simulation with the second-order (top panel) and  fourth-order (bottom panel)
Trotter product formulas and the corresponding MPFs.
The cost is measured by the number of Trotter steps required to approximate the time-evolved
state $\rho(t)$ within 1\% approximation error.
MPFs parameters were chosen according to Fig.~\ref{fig:table_examples}.}
     \label{fig:one_percent_sim}
\end{figure}

The error tolerance $10^{-3}$ might be too ambitious goal for  near-term
quantum processors  lacking error correction capabilities.
Consider a more realistic goal of simulating $n$-qubit Hamiltonian Eq.~(\ref{spin_chain})
over time interval $t=n$ within  1\% error (with respect to the trace norm distance).
Combining  fitting formulas Eqs.~(\ref{order2fit},\ref{order2fit_Tro},\ref{order4fit},\ref{order4fit_Tro})
we can compare the number of Trotter steps required to solve this simulation task
using (a) second-order and (b) fourth-order  Trotter product formulas and their multi-product version.
This comparison is presented on Fig.~\ref{fig:one_percent_sim}.

\paragraph{Minimax MPFs.}
Here we demonstrate the performance of Algorithm~\ref{alg:mdMPF} with
noisy Gram matrix $M$ and approximated data $L^{\apx}$ and compare it against the dynamic MPF with exact data ($M$ and $L^\exact$), static MPF and the best Trotter product formula within MPF. To this end we use the same Hamiltonian as in the previous section, but employ 5 Trotter circuits $\rho_{k_i}$ of 2nd order with $\vec k_5=(k_1,k_2,k_3,k_4,k_5)^\top=2\cdot(4,10, 13,15,17)^\top$. This choice is made to stress one practically important advantage of minimax MPF over the static MPF: the former can provide coefficients of low condition number for any sequence of $k_i$ in contrast to the static MPF, which is sensitive to the choice of $k_i$: indeed, for the above chosen $k_i$ the resulting Vandermonde matrix~\eqref{ZNE} is ill-conditioned numerically, hence the vector of coefficients resulting from solving~\eqref{ZNE}, $\vec c^{(5)}=(8.93717538e-05, -7.30297487e-01,  1.14983897e+01, -2.73719522e+01, 1.76037706e+01)^\top$ has large condition number, $\|\vec c^{(5)}\|_1\approx 57$. Clearly, $\vec c^{(5)}$ is impractical since its last three components will amplify sampling noise when computing the linear combination $\sum_{i=1}^5 c^{(5)}_i \tr(O\rho_{k_i}(t))$ provided $\tr(O\rho_{k_i}(t))$ is computed on a noisy quantum hardware. However, as noted in Fig.~\ref{fig:table_examples}, for certain $k_i$ the static MPF formula can be made well-coditioned: e.g. picking a subset of $\vec k_5$, namely $\vec k_3=2\cdot (4,13,17)^\top$ one can construct a well-conditioned static MPF with coefficients $\vec c^{(3)}=(0.00612895, -1.55561002, 2.54948107)^\top$ and condition number $\|\vec c^{(3)}\|_1\approx 5$. Below we compare performance of this well-conditioned static MPF given by $\sum_{i=1,k_i\in\{8,26,34\}}^3 c^{(3)}_i \rho_{k_i}$ versus dynamic and minimax MPFs corresponding to $\vec k_5$: this comparison will make it apparent that minimax MPF allows one to add more of ``short depth'' Trotter circuits to the linear combination, namely $\rho_{20}$ and $\rho_{30}$ in this case, resulting in further improvement of the precision compared to the well-conditioned static MPF with three circuits $\rho_{8,26,34}$ without increasing the condition number. 


As input for Algorithm~\ref{alg:mdMPF} we use:
\begin{itemize}
    \item $t_0=1$, $dt=0.05$, $\oc(t_0)=(0.00612895, 0, -1.55561002, 0,  2.54948107)^\top$
    \item final time $T$ is such that the Frobenious error of the dynamic MPF with exact data is below $0.31$,
    \item $\varepsilon=0.01$ and $\bar M(t)=M(t)+E_1^D$, $\bar A(t)=Q(t)+E_2^D$ where $(i,j)$-component of $M$, $Q$ is computed as per~\cref{eq:Mt,eq:Qt} (we use $k_0=26$ in~\cref{eq:Qt}), and $E_{1,2}^D$ are drawn from normal distribution for every time $t$ such that $\|E_{1,2}^D\|_{2,2}\le\varepsilon$ and the components of $\bar M$, $\bar A$ are non-negative, $\bar M$ has ones on its diagonal.
\end{itemize}
To compute the coefficients of the dynamic MPF with exact data we minimize quadratic form~\eqref{proj_error} over $\vec c: \mathbb{1}^\top \vec c=1$ for every $t_j=t_0+j dt$. The dimension of this convex optimization problem is very small: $\vec c$ has only 5 components; the same observation applies to the convex optimization problem of Algorithm~\ref{alg:mdMPF}. Having this in mind, we solve all those optimization problems by employing Python package cvxpy~\cite{diamond2016cvxpy}. Corresponding snippets of Python code are provided in the Appendix.

Figure~\ref{plot:dMPF} provides the comparison: shortly before time $t=2$ minimax MPF outperforms the best Trotter product formula ($k=34$), it gets $10\times$ better for $t=2.8$, and around time $t=3.9$ the minimax MPF outperforms the static MPF, it gets $5\times$ better for $t=4.5$, namely when the error of minimax MPF reaches the shot noise magnitude, with condition number $\|\hat c(4.5)\|_1\approx 5$. Clearly, dynamic MPF with exact data delivers the best performance: about $10000\times$ better than the static MPF and $100000\times$ better than the Trotter formula for $t=3.29$. However its condition number is quite high: $\|\oc(3.29)\|_1\approx 46$ in contrast to $\|\vec c_3\|_1\approx 5$ of static MPF and to $\|\hat c(3.29)\|_1=2.17$ of the minimax MPF. The difference in performance between dynamic MPF and minimax MPF in the presence of noise is due to the penalization term $\varepsilon \|x\|_2$ in Algorithm~\ref{alg:mdMPF}. This term imposes extra constraints on the coefficients, and acts to simultaneously reduce the effect of sampling noise and the condition number of the coefficients. In the presence of shot noise, minimax MPF is therefore more practical than dynamic MPF.

\begin{figure}[h]
     \centering
     \includegraphics[width=9cm]{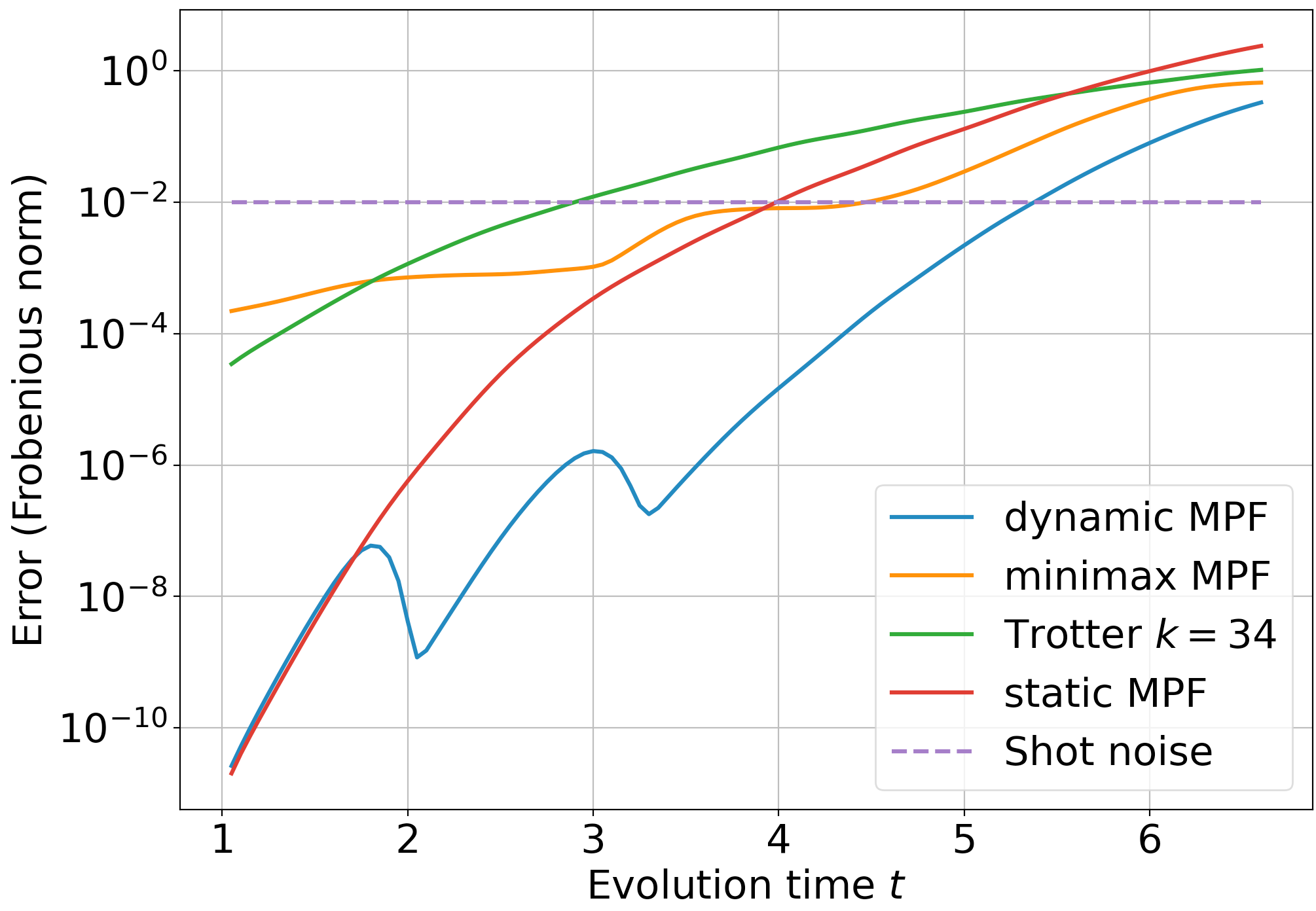}
     \caption{Approximation error for MPFs: well-conditioned static MPF with $\rho_{8,26,34}$ (red), best Trotter formula $\rho_{34}$ (green), dynamic MPF with exact data (blue) and minimax MPF with noisy/approximated data (orange) both using $\rho_{8,20, 26,30, 34}$, for the Heisenberg spin chain Hamiltonian Eq.~(\ref{spin_chain}) with $n=10$ qubits. The shot noise magnitude used to generate $\bar M,\bar A$ is depicted as a dashed line for reference.
     \label{plot:dMPF}}
\end{figure}

\section{Conclusions}

To conclude, we studied quantum algorithms for simulating Hamiltonian dynamics based
on multi-product formulas. Our first contribution, the upper bound on the approximation error achieved
by MPF algorithms of Trotter type for general quantum Hamiltonians which are commonly used in quantum simulations.
This upper bound applies to both short and long evolution times overcoming limitations of
the previously known bounds. Numerical simulations confirm that our upper bound correctly predicts the
approximation error scaling for the considered spin chain Hamiltonian.

Our second contribution, minimax multi-product formula with time-dependent coefficients minimizes Trotter error in Frobenious norm. Minimax MPF is designed to be robust to algorithmic
and hardware errors, namely to small additive error in the estimation
of overlaps between Trotter circuits which are unavoidable in
the experimental implementation due to sampling and hardware noise. Even in the presence of noise minimax MPF outperforms both well-conditioned MPF and the best Trotter formula in numerical simulations. We prove that the worst-case error of minimax MPF scales
proportionally to the upper bound of~\Cref{thm:MPF}.

A challenging open question is how to extend our bound, Theorem~\ref{thm:MPF},
to the case $r>p+1$ and obtain a stronger than quadratic error suppression (in the worst case) compared
with the regular product formulas. Likewise, one may ask whether the upper bound of Theorem~\ref{thm:MPF}
can be expressed only in terms of low-order nested commutators similar to
the bound of Lemma~\ref{fact:trotter_error}.

\section*{Acknowledgements}
SB thanks  Minh Tran, Almudena Vazquez, and
 Stefan W\" orner for  helpful discussions.
SB was supported in part by the Army Research Office under Grant Number W911NF-20-1-0014.

\appendix
\onecolumngrid

\section{Trotter error for Multi Product Formulas}
\label{app:A}

\subsection*{Proof of Lemma~\ref{fact:trotter_error}}

Consider first a single time step, $k=1$.
Then our goal is to upper bound
the error  $\|\rho_1(t)-\rho(t)\|_1$ where
\[
\rho(t)=e^{-itH}\rho_{in} e^{itH} \quad \mbox{and} \quad
\rho_1(t)=\calS(t) \rho_{in} \calS(t)^\dag.
\]
Recall that   $\calS(t)=e^{-itF_d} \cdots e^{-itF_1}$ is an order-$p$ product formula
associated with a Hamiltonian $H=\sum_{a=1}^d F_a$.
Time evolution of $\calS(t)$ is governed by
a time dependent Hamiltonian $G(t)$ defined as
\be
\label{dotS}
G(t) =  i\left( \frac{d\calS(t)}{dt}\right) \calS(t)^{-1}.
\ee
The chain rule for derivatives gives
\be
\label{G(t)}
G(t)= F_d + \sum_{a=2}^{d} e^{-itF_d}  \cdots e^{-itF_{a}}F_{a-1} e^{itF_{a}}\cdots e^{itF_d}.
\ee
Note that $G(0)=\sum_{a=1}^d F_a = H$.
A simple algebra shows that
\[
\frac{d}{dt}(\rho_1(t) - \rho(t))=-i[G(t),\rho_1(t)]+i[H,\rho(t)]
=-i[H,\rho_1(t)-\rho(t)] -i[G(t)-H,\rho_1(t)].
\]
Here we have added and subtracted a term $i[H,\rho_1(t)]$ to get the second equality.
The above equation is equivalent to
\[
\frac{d}{dt} \left( e^{iHt} (\rho_1(t)-\rho(t))e^{-iHt}\right)
=-ie^{itH} [G(t)-H,\rho_1(t)]e^{-itH}.
\]
Integrating over the interval $[0,t]$ and noting that $\rho_1(0)=\rho(0)$
gives
\be
\label{error_as_integral}
e^{iHt} (\rho_1(t)-\rho(t))e^{-iHt} = -i\int_0^t dt_1 e^{it_1H} [G(t_1)-H,\rho_1(t)]e^{-it_1H}.
\ee
Using the unitarity of $e^{itH}$ and  the triangle inequality one obtains
\be
\label{triangle_ineq_bound}
\| \rho_1(t)-\rho(t)\|_1 \le \int_0^t dt_1 \| \,[G(t_1)-H,\rho_1(t_1)]\, \|_1
\le 2  \int_0^t dt_1 \| G(t_1) -H\|.
\ee
The second inequality relies on the bound $\|XY\|_1\le \|X\|\cdot \|Y\|_1$
which holds for any operators $X,Y$ and the identity $\|\rho_1(t_1)\|_1=1$.
We shall need
the following  result which is a rephrasing of Theorem~5 from Ref.~\cite{childs2021theory}.
\begin{fact}
\label{fact:remainder}
Suppose $A_1,\ldots,A_s,B$ are hermitian operators.
Consider an operator-valued function
\[
B(t) =e^{-itA_1} \cdots e^{-itA_s} B e^{itA_s} \cdots e^{itA_1}
\]
and its Taylor series $B(t)=\sum_{q=0}^\infty B_q t^q$ at $t=0$. Then
\be
\left\| B(t) - \sum_{q=0}^{p-1} B_q t^q\right\| \le \alpha_{\comm}(p;A_1,\ldots,A_s;B) \frac{|t|^p}{p!}
\ee
where $\alpha_{\comm}(p;A_1,\ldots,A_s;B)$ is defined in Eq.~(\ref{alpha_comm}).
\end{fact}
Consider Taylor series
\[
G(t)=\sum_{q=0}^\infty G_q t^q.
\]
We claim that
\be
\label{Taylor_coeff}
G_0=H \quad \mbox{and} \quad G_q=0 \quad \mbox{for $1\le q\le p-1$}.
\ee
Indeed, Eq.~(\ref{G(t)}) gives $G_0=G(0) = \sum_{a=1}^d F_a=H$.
The assumption that $\calS(t)$ is an order-$p$ product formula
implies that $\rho_1(t)-\rho(t)=O(t^{p+1})$ in the limit $t\to 0$,
see Eq.~(\ref{1norm2operator}).
From Eq.~(\ref{triangle_ineq_bound}) one gets
$G(t)-H=O(t^p)$ in the limit $t\to 0$.
Thus $G_q=0$ for $1\le q\le p-1$.
Combining Eqs.~(\ref{G(t)},\ref{Taylor_coeff}) and Fact~\ref{fact:remainder} gives
\be
\label{calT_upper_bound}
\|G(t)-H\|
=\left\| G(t)-\sum_{q=0}^{p-1} G_q t^q\right\|
\le \frac{2t^p}{p!}  \sum_{a=2}^{d} \alpha_{\comm}(p;F_d,\ldots,F_{a};F_{a-1})
= \frac{2\alpha_p t^p}{p!}.
\ee
Here we applied Fact~\ref{fact:remainder} to each term
$e^{-itF_d}  \cdots e^{-itF_{a}}F_{a-1} e^{itF_{a}}\cdots e^{itF_d}$
in $G(t)$ with $a=2,\ldots,d$ and used the triangle inequality.
Substituting this into Eq.~(\ref{triangle_ineq_bound}) gives
\[
\| \rho_1(t)-\rho(t)\|_1\le 2  \int_0^t dt_1 \frac{2\alpha_p t_1^p}{p!}
= \frac{2\alpha_p t^{p+1}}{(p+1)!}.
\]
This is equivalent to Eq.~(\ref{childs_su}) with a single time step $k=1$.

Consider now the general case $k\ge 1$.
Define a sequence of states
\[
\sigma_j = \calS(t/k)^{k-j} e^{-itH(j/k)}\rho_{in} e^{itH(j/k)}  \calS(t/k)^{-k+j}
\]
where $j=0,1,\ldots,k$.  We have $\sigma_0=\rho_k(t)$ and $\sigma_k=\rho(t)$.
The norm $\|\sigma_{j+1} -\sigma_{j}\|_1$ can be bounded using Eq.~(\ref{childs_su}) with $k=1$,
evolution time $t/k$, and the initial state $e^{-itH(j/k)}\rho_{in} e^{itH(j/k)}$.
By the triangle inequality,
\[
\|\rho_k(t)-\rho(t)\|_1 = \| \sigma_0 - \sigma_k\|_1 \le \sum_{j=0}^{k-1} \| \sigma_{j+1} -\sigma_j\|_1 \le
 \frac{2k\alpha_p (t/k)^{p+1}}{(p+1)! }=
   \frac{2\alpha_p t^{p+1}}{(p+1)! k^p}.
\]
This proves Eq.~(\ref{childs_su}) in the general case.

\subsection*{Proof of Theorem~\ref{thm:MPF}}

Below we use notations from the previous subsection.
Fix evolution time $t>0$ and the number of time steps $k$.
Any $\tau\in [0,t)$ can be uniquely written as
\be
\tau = \tau' +  (t/k)j(\tau) \quad \mbox{for  $\tau'\in [0,t/k)$ and integer $j(\tau)\in [0,k-1]$}.
\ee
Define an operator valued function
\be
\label{Floq1}
\calU(\tau) =  \calS(\tau')\calS(t/k)^{j(\tau)}.
\ee
By definition,
$\calU(0)=I$ and $\lim_{\tau\to t} \calU(\tau)=\calS(t/k)^k$. Furthermore,
for any $\tau \in [0,t)$ which is not an integer multiple of $t/k$
the function $\calU(\tau)$ is differentiable and
\be
\label{Floq2}
i\frac{d \calU(\tau)}{d\tau} = i\frac{d\calS(\tau')}{d\tau'} \calS(t/k)^{j(\tau)}
=G(\tau') \calS(\tau')  \calS(t/k)^{j(\tau)}  =  G(\tau') \calU(\tau).
\ee
Here the second equality uses Eq.~(\ref{dotS}).
Consider a state
\be
\sigma(\tau) = \calU(\tau) \rho_{in} \calU(\tau)^\dag, \qquad 0\le \tau < t.
\ee
By definition,
\be
\lim_{\tau \to t} \sigma(\tau) = \calS(t/k)^k \rho_{in} \calS(t/k)^{-k} = \rho_k(t).
\ee
Repeating the same arguments as in the derivation of Eq.~(\ref{error_as_integral})
one gets
\be
\label{integral_first_order}
e^{i\tau \ad_H} (\sigma(\tau)-\rho(\tau))= -i\int_0^\tau d\tau_1 e^{i\tau_1 \ad_H} [G(\tau_1')-H,\sigma(\tau_1)].
\ee
Adding and subtracting $[G(\tau_1')-H,\rho(\tau_1)]$ in the righthand side
gives
\be
\label{integral_first_order_1}
e^{i\tau \ad_H} (\sigma(\tau)-\rho(\tau)) =
J_1(\tau)  -i\int_0^\tau d\tau_1 e^{i\tau_1\ad_H} [G(\tau_1')-H,\sigma(\tau_1)-\rho(\tau_1)],
\ee
where
\be
J_1(\tau) =-i\int_0^\tau d\tau_1 e^{i\tau_1\ad_H} [G(\tau_1')-H,\rho(\tau_1)]
\ee
We can now apply Eq.~(\ref{integral_first_order}) again to express
$\sigma(\tau_1)-\rho(\tau_1)$ in Eq.~(\ref{integral_first_order_1}) as an integral.
This gives
\be
\label{sigma_rho_1}
e^{i\tau\ad_H} (\sigma(\tau)-\rho(\tau)) = J_1(\tau) + J_2(\tau),
\ee
where
\be
J_2(\tau) =
-\int_0^\tau d\tau_1 \int_0^{\tau_1} d\tau_2
[ e^{i\tau_1 \ad_H} (G(\tau_1')-H),e^{i\tau_2 \ad_H}[G(\tau_2')-H,\sigma(\tau_2)]].
\ee
Using  the triangle inequality and normalization $\|\sigma(\tau_2)\|_1=1$ one gets
\be
\|J_2(\tau)\|_1 \le 4\int_0^\tau d\tau_1 \int_0^{\tau_1} d\tau_2
\| G(\tau_1')-H\| \cdot \| G(\tau_2')-H\|.
\ee
Consider the limit $\tau \to t$.
Dividing the integration domain into $k$ regions with $j(\tau_1)=j(\tau_2)$
and ${k \choose 2}$ regions with $j(\tau_1)>j(\tau_2)$ leads to
\be
\|J_2(t)\|_1 \le
4k \int_0^{t/k} d\tau_1' \int_0^{\tau_1} d\tau_2' \| G(\tau_1')-H\| \cdot \| G(\tau_2')-H\|+
4 {k \choose 2} \left( \int_0^{t/k} d\tau_1' \| G(\tau_1') - H\| \right)^2.
\ee
From Eq.~(\ref{calT_upper_bound}) one gets $\|G(\tau_1)-H\|\le 2\alpha_p \tau_1^p/p!$
and $\|G(\tau_2)-H\|\le 2\alpha_p \tau_2^p/p!$.
Taking the integrals gives
\be
\label{sigma_rho_2}
\|J_2(t)\|_1 \le  (2k + 2k(k-1)) \left( \frac{2\alpha_p t^{p+1}}{(p+1)! k^{p+1}}\right)^2
=2 \left( \frac{2\alpha_p t^{p+1}}{(p+1)! k^p}\right)^2.
\ee
Note that $\|J_2(t)\|_1$ is proportional to the square of the regular Trotter error,
see Eq.~(\ref{childs_su}).

Next let us examine the term $J_1(t)$. By definition, $\rho(\tau_1)=e^{-i\tau_1 \ad_H} \rho_{in}$.
Thus
\be
\label{sigma_rho_3}
J_1(t) = -i [L(t),\rho_{in}]
\ee
where
\be
\label{L1}
L(t) = \int_0^t d\tau_1 e^{i\tau_1 \ad_H} (G(\tau_1') - H).
\ee
Writing $\tau_1 = (t/k)(v+j)$ with $v\in [0,1)$ and integer $j\in [0,k-1]$ we get
\be
\label{L2}
L(t) =\frac{t}{k} \sum_{j=0}^{k-1} \int_0^1 dv
e^{i j(t/k) \ad_H }e^{i v(t/k) \ad_H}(G(vt/k) - H).
\ee
The dependence of $L(t)$ on $1/k$ is hard to analyze because $k$ controls the range
of the sum $\sum_{j=0}^{k-1}$. On the other hand, one should expect
that the sum $\sum_{j=0}^{k-1}$ can be well approximated by an integral $\int_0^k dx$. The latter can be transformed
to an integral over a range $[0,1]$ independent of $k$ by a simple change of variable.
A systematic way to approximate a sum by an integral is based
on  the Euler–Maclaurin formula, see e.g. Theorem~D.2.1 in~\cite{andrews_askey_roy_1999}.
\begin{lemma}[\bf Euler–Maclaurin formula]
Suppose $f(x)$ is function with continuous derivatives up to order $s$.
Let $B_s(x)$ be the Bernoulli polynomial of order $s$ and $B_s=B_s(0)$ be
the Bernoulli  number. For example, $B_0(x)\equiv 1$, $B_1(x)=x-1/2$,
and $B_2(x)=x^2 -x + 1/6$.
Then for any integers $j_1<j_2$ one has
\be
\label{Euler}
\sum_{j=j_1+1}^{j_2} f(j) = \int_{j_1}^{j_2} dx f(x) + \sum_{\ell=1}^{s}(-1)^{\ell} \frac{B_\ell}{\ell !}
\left( f^{(\ell-1)}(j_2) - f^{(\ell-1)}(j_1)\right)+
\frac{(-1)^{s-1}}{s!} \int_{j_1}^{j_2} dx B_s(x-[x]) f^{(s)}(x).
\ee
Here $f^{(\ell)}(x)$ denotes the $\ell$-th derivative of $f(x)$ and $[x]$ denotes the integer part of $x$
such that $x-[x]$ always lies in the interval $[0,1)$.
Furthermore, for all $s\ge 2$
\be
\label{Bernoulli_upper}
\max_{x\in [0,1]}\;
|B_s(x)| \le \frac{4s!}{(2\pi)^s}.
\ee
\end{lemma}
Consider a  function $f(x)=e^{i (x-1)(t/k)\ad_H}$.
Note that $x$ is a real variable while $f(x)$ is a super-operator
(a linear map acting on the space of linear operators).
We have
\[
f^{(\ell)}(x) = (it/k)^\ell (\ad_H)^\ell f(x).
\]
Applying Eq.~(\ref{Euler})  with $j_1=0$ and
$j_2=k$ gives
\begin{align}
\sum_{j=0}^{k-1} e^{i j(t/k) \ad_H } &= \sum_{j=1}^k f(j)
=  \int_{0}^k dx e^{i(x-1)(t/k) \ad_H}  \\
&
+ \sum_{\ell=1}^s (-1)^{\ell}\frac{B_\ell (it)^{\ell-1}}{k^{\ell-1} \, \ell !}  \left( e^{it \ad_H} - \hat{I}\right)
e^{-i(t/k) \ad_H}
(\ad_H)^{\ell-1}
\nonumber \\
&
+\frac{(-1)^{s-1}(it)^s}{k^s \, s!} \int_0^k dx \, B_s(x-[x])e^{i (x-1)(t/k)\ad_H}(\ad_H)^s.
\end{align}
Here $\hat{I}$ is the identity superoperator.
Performing a change of variables $x=kw$ with $w\in [0,1]$ in the integral over $x$ and
inserting the above expression into Eq.~(\ref{L2}) gives
\begin{align}
L(t) & = t \int_0^1 dw e^{itw \ad_H}  \int_0^1 dv\, e^{i(v-1)(t/k) \ad_H}(G(tv/k) - H) \nonumber \\
&
- \left( e^{it \ad_H} - \hat{I}\right)\sum_{\ell=1}^s (-i)^{\ell}\frac{B_\ell t^{\ell} }{\ell !}
\int_0^1 dv\,
\frac1{k^\ell}(\ad_H)^{\ell-1}
e^{i(v-1)(t/k) \ad_H}(G(tv/k) - H)
\nonumber\\
&
-\frac{(-i)^{s}t^{s+1}}{k^s \, s!} \int_0^1 dw e^{itw \ad_H} B_s(kw-[kw]) \int_0^1 dv \,
(\ad_H)^{s}e^{i(v-1)(t/k) \ad_H}(G(tv/k) - H).
\label{sigma_rho_4}
\end{align}
Combining Eqs.~(\ref{sigma_rho_1},\ref{sigma_rho_2},\ref{sigma_rho_3},\ref{Bernoulli_upper},\ref{sigma_rho_4})
and using the triangle inequality gives
\be
\label{e1e2e3}
\left\| \mu(t)-\rho(t) \right\|_1 \le 2(\epsilon_1+\epsilon_2 +\epsilon_3) + 2 \left( \frac{2\alpha_p t^{p+1}}{(p+1)! }\right)^2
\sum_{i=1}^r \frac{|c_i|}{k_i^{2p}}
\ee
where
\be
\epsilon_1 = t\max_{v \in [0,1]}\;  \left\| \sum_{j=1}^r c_j e^{i(v-1)(t/k_j) \ad_H}(G(tv/k_j)-H)\right\|,
\ee
\be
\epsilon_2 = \sum_{\ell=1}^s  \frac{2B_\ell t^{\ell}}{ \ell !}
\max_{v \in [0,1]}\;
 \left\|
\sum_{j=1}^r \frac{c_j}{k_j^\ell} (\ad_H)^{\ell-1}e^{i(v-1)(t/k_j) \ad_H}(G(tv/k_j)-H)
\right\|
\ee
and
\be
\epsilon_3 = \frac{4t^{s+1}}{(2\pi)^s} \max_{v \in [0,1]}\;
\sum_{j=1}^r \frac{|c_j|}{k_j^s}  \cdot \| (\ad_H)^s e^{i(v-1)(t/k_j) \ad_H} (G(tv/k_j)-H)\|.
\ee
Below we choose
\be
r=p+1 \quad \mbox{and} \quad s=p.
\ee
In order to bound $\epsilon_1$ and $\epsilon_2$ we have to use condition Eq.~(\ref{ZNE}), that is,
\be
\label{ZNErestated}
\sum_{j=1}^r c_j =1
\quad \mbox{and} \quad
\sum_{j=1}^r \frac{c_j}{k_j^q} = 0
\quad \mbox{for $q\in \{p,p+1,\ldots,p+r-2\}$}.
\ee
Consider the term $\epsilon_1$.
 Using Eq.~(\ref{ZNErestated}) one can
replace $e^{i(v-1)(t/k_j) \ad_H}(G(tv/k_j)-H)$ in $\epsilon_1$ by its Taylor series with all terms
of order less than $p+r-1=2p$ omitted. Here the Taylor series are computed with respect
to the variable $t/k_j$.
The norm of such truncated Taylor
series can be bounded using Fact~\ref{fact:remainder}. It gives
\be
\epsilon_1 \le
\frac{2 t^{2p+1}}{(2p)!} \left(\sum_{i=1}^{p+1}  \frac{|c_i|}{k_i^{2p}}\right)
 \sum_{a=2}^d \alpha_{\comm}(2p;H,F_d,\ldots, F_a; F_{a-1}).
\ee
To bound $\epsilon_2$ and $\epsilon_3$ we shall need the following generalization
of Fact~\ref{fact:remainder}.
\begin{lemma}
\label{lemma:remainder}
Suppose $A_1,\ldots,A_s,B$ are hermitian operators
and $\ell\ge 0$ is an integer.
Consider an operator-valued function
\[
B(t) =(\ad_{A_1})^{\ell} e^{-it\ad_{A_1}} e^{-it\ad_{A_2}} \cdots  e^{-it\ad_{A_s}} (B)
\]
and its Taylor series $B(t)=\sum_{q=0}^\infty B_q t^q$ at $t=0$.
Define a set of unitary operators
\[
\Gamma(t) = \{ e^{-i\tau_2 A_2} \cdots e^{-i\tau_s A_s} \, \vert \,
0\le \tau_2,\ldots, \tau_s\le t\}.
\]
Then for all $t\ge 0$
\be
\label{fact1_generalized}
\left\| B(t) - \sum_{q=0}^{p-1} B_q t^q\right\| \le
\frac{t^p}{p!}\alpha_{\comm}(p,\ell;A_1,\ldots,A_s;B;t),
\ee
with
\be
\alpha_{\comm}(p,\ell;A_1,\ldots,A_s;B;t)=
 \sum_{\substack{j_1,\ldots,j_s\ge 0\\ j_1+\ldots+j_s=p\\}}\;\;
\frac{p!}{j_1! \cdots j_s!} \; \;
\max_{U\in \Gamma(t)}
\left\|
(\ad_{A_1})^{\ell+j_1} \hat{U} \prod_{\gamma=2}^s (\ad_{A_\gamma})^{j_\gamma} (B)
\right\|. \nonumber
\ee
Here $\hat{U}$ is a linear map such that $\hat{U}(X)=UXU^{-1}$ for all operators $X$.
\end{lemma}
We postpone the proof of the lemma until the end of this section.
Consider the term $\epsilon_2$.
Using Eq.~(\ref{ZNErestated}) one can
replace the operator
$(\ad_H)^{\ell-1} e^{i(v-1)(t/k_j) \ad_H}(G(tv/k_j)-H)$ in $\epsilon_2$ by its Taylor series with all terms
of order less than $p+r-1-\ell=2p-\ell$ omitted. The norm of such truncated Taylor
series can be bounded using Lemma~\ref{lemma:remainder}.
It gives
\be
\epsilon_2 \le \sum_{\ell=1}^p  \frac{2B_\ell   t^{2p}}{\ell! (2p-\ell)!}
\left(
\sum_{j=1}^{p+1} \frac{|c_j|}{k_j^{2p}}\right) \sum_{a=2}^d
\alpha_{\comm}(2p-\ell,\ell-1;H,F_d,\ldots, F_a; F_{a-1};t/k_{min})
\ee
Consider the term $\epsilon_3$ with $s=p$.
The assumption that the base product formula has order $p$ implies
that one can replace the operator
$ (\ad_H)^p e^{i(v-1)(t/k_j) \ad_H} (G(tv/k_j)-H)$ in $\epsilon_3$
by its Taylor series with all terms of order less than $p$ omitted.
The norm of such truncated Taylor
series can be bounded using Lemma~\ref{lemma:remainder}.
It gives
\be
\epsilon_3 \le  \frac{4 t^{2p+1}}{(2\pi)^p p!}\left( \sum_{j=1}^{p+1} \frac{|c_j|}{k_j^{2p}}\right)
 \sum_{a=2}^d
\alpha_{\comm}(p,p;H,F_d,\ldots, F_a; F_{a-1};t/k_{min}).
\ee
Substituting the bounds on $\epsilon_1,\epsilon_2,\epsilon_3$
into Eq.~(\ref{e1e2e3})
and using the definition of commutator norms $\beta_{p_1,p_2}$, see Eq.~(\ref{beta_comm1},\ref{beta_comm2}),
completes the proof of Theorem~\ref{thm:MPF}.

\subsection*{Proof of Lemma~\ref{lemma:remainder}}

First let us record the well-known integral representation of truncated Taylor series of the exponential function.
\begin{lemma}[\bf Taylor's theorem]
\label{lemma:exp_int}
For any operator $X$, evolution time $t\ge 0$, and integer $m\ge 1$ one has
\be
\sum_{q=m}^\infty \frac1{q!} (-it\ad_{X})^q
=
\left( \int_{0}^t d\tau  \mu(\tau)e^{-i\tau \ad_{X}} \right)\frac{(-it\ad_{X})^m}{m!}
\ee
where $\mu(\tau)$ is a function satisfying $\mu(\tau)\ge 0$ and
$\int_0^t d\tau \mu(\tau)=1$.
\end{lemma}
\begin{proof}
Applying Taylor's theorem with integral form of the remainder
to the exponential function $e^{-it\ad_X}$
gives
\[
\sum_{q=m}^\infty \frac1{q!} (-it\ad_X)^q
=\left( mt^{-m} \int_0^t d\tau (t-\tau)^{m-1}e^{-i\tau \ad_X}\right)
\frac{(-it\ad_{X})^m}{m!}
\]
This is the desired representation with $\mu(\tau) = mt^{-m} (t-\tau)^{m-1}$.
\end{proof}
\begin{proof}[\bf Proof of Lemma~\ref{lemma:remainder}]
First let us introduce some notations.
Below we consider  $s$-tuples of non-negative integers $J=(j_1,\ldots,j_s)$
satisfying $\sum_{\gamma=1}^s j_\gamma=p$. Let $\Omega_p$ be the set of
all such tuples. For any $J\in \Omega_p$
let $m(J)$ be largest integer in the range $\{1,2,\ldots,s\}$ such that
$\sum_{\gamma =m(J)}^ s j_\gamma \ge p$.
Note that $j_{m(J)}\ge 1$ and $\sum_{\gamma=m(J)+1}^s j_\gamma \le p-1$.
We use a shorthand $J!\equiv \prod_{\gamma=1}^s (j_\gamma)!$.

We need an upper bound on the norm of
the remainder operator $R(t)= B(t) - \sum_{q=0}^{p-1} B_q t^q$.
 Expanding each exponential function that appears in $B(t)$ in Taylor series
gives
\[
R(t)=\sum_{m=1}^s R_m(t),
\]
where
\[
R_m(t) = (\ad_{A_1})^{\ell}  \sum_{J\in \Omega_p\, : \, m(J)=m}\;  \frac1{J!}  (-it\ad_{A_1})^{j_1} (-it\ad_{A_2})^{j_2}
\cdots  (-it\ad_{A_s})^{j_s}(B).
\]
The sum over $J\in \Omega_{p}$ with $m(J)=m$ can be rewritten as
\[
 \sum_{J\in \Omega_{p}\, : \, m(J)=m} = \sum_{j_1=0}^\infty \cdots \sum_{j_{m-1}=0}^\infty\;
 \sum_{\substack{j_{m+1},\ldots j_s \ge 0\\ j_{m+1} +\ldots+j_s\le p-1\\}}\;
 \sum_{j_m=p-j_{m+1}-\ldots-j_s}^{\infty}.
\]
Thus
\[
R_m(t)= (\ad_{A_1})^{\ell} e^{-it\ad_{A_1}} \cdots  e^{-it\ad_{A_{m-1}}}
\sum_{\substack{j_{m+1},\ldots j_s \ge 0\\ j_{m+1} +\ldots+j_s\le p-1\\}}\;
\sum_{j_m=p-j_{m+1}-\ldots-j_s}^{\infty}
\frac1{j_m! \cdots j_s!}
 (-it\ad_{A_m})^{j_m}(C_{j_{m+1},\ldots,j_{s}})
\]
where
\[
C_{j_{m+1},\ldots,j_{s}}=   (-it\ad_{A_{m+1}})^{j_{m+1}}\cdots (-it\ad_{A_s})^{j_s}(B).
 \]
Applying Lemma~\ref{lemma:exp_int} to compute the sum over $j_m$ gives
\[
\sum_{j_m=p-j_{m+1}-\ldots-j_s}^{\infty} \; \;  \frac1{j_m!}
 (-it\ad_{A_m})^{j_m}
=\int_0^t  d\tau \mu(\tau)  e^{-i\tau \ad_{A_m}} \frac{(-it\ad_{A_m})^{p-j_{m+1}-\ldots-j_s}}{(p-j_{m+1}-\ldots-j_s)!},
\]
where $\mu(\tau)\ge 0$ is a normalized probability measure.
Plugging this into the  formula  for $R_m(t)$ one gets
\[
R_m(t) =(-it)^p \int_0^t d\tau \mu(\tau)(\ad_{A_1})^{\ell} e^{-it\ad_{A_1}} \cdots  e^{-it\ad_{A_{m-1}}} e^{-i\tau \ad_{A_m}}
\sum_{\substack{j_1,\ldots,j_s\ge 0\\
j_1+\ldots+j_s=p\\
j_1=\ldots=j_{m-1}=0\\
 j_m\ge 1\\}}\; \;  \frac1{J!} (\ad_{A_m})^{j_m} \cdots (\ad_{A_s})^{j_s}(B).
\]
Since $e^{-it\ad_{A_1}}$ commutes with $(\ad_{A_1})^\ell$
and $\| e^{-it\ad_{A_1}}(X)\|=\|X\|$ for any operator $X$, we get
\[
\|R_m(t)| \le t^p
\sum_{\substack{j_1,\ldots,j_s\ge 0\\
j_1+\ldots+j_s=p\\
j_1=\ldots=j_{m-1}=0\\
 j_m\ge 1\\}}\; \; \frac1{J!}
\max_{U \in \Gamma(t)}\;
\left\| (\ad_{A_1})^{\ell+j_1} \hat{U}
 (\ad_{A_2})^{j_2} \cdots (\ad_{A_s})^{j_s}(B)\right\|.
 \]
The triangle inequality $\|R(t)\|\le \sum_{m=1}^s \|R_m(t)\|$ then gives the desired bound.
\end{proof}

\section{MPF Trotter error for $k$-local spin Hamiltonians}
\label{app:linear_scaling_with_n}

In this section we examine the coefficients $a_2$ and $a_3$
that appear in the upper bound of Theorem~\ref{thm:MPF}.
We specialize the theorem to Hamiltonians
$H$ describing a system of $n$ qubits with $k$-qubit interactions
known as $k$-local Hamiltonians~\cite{kitaev2002classical}.
More precisely, let  $\calL(k,J)$ be the set of all $n$-qubit operators $O$
that admit a decomposition $O=\sum_\alpha O_\alpha$
such that each term $O_\alpha$ acts non-trivially on at most $k$
qubits and
\[
\sum_{\alpha \, : \, \mathrm{supp}(O_\alpha)\ni q} \; \;
\| O_\alpha\| \le J
\]
for any qubit $q$.
Here $\mathrm{supp}(O_\alpha)$ is the support of $O_\alpha$,
that is, the subset of qubits where $O_\alpha$ acts non-trivially.
Thus the above sum includes all terms $O_\alpha$ that act non-trivially
on a given qubit $q$.
Hermitian operators in $\calL(k,J)$ are $k$-local
Hamiltonians with an interaction strength $J$.
For example, any Hamiltonian describing a $D$-dimensional cubic lattice of qubits with bounded-norm interactions supported on elementary cubes is contained in $\calL(2^D,J)$
for some constant $J=O(1)$.
Note that any operator $O\in \calL(k,J)$ obeys
\[
\|O\|\le \sum_\alpha \|O_\alpha\| \le \sum_{q=1}^n \; \sum_{\alpha \, : \, \mathrm{supp}(O_\alpha)\ni q} \; \;
\| O_\alpha\| \le nJ.
\]

Consider an order-$p$ product formula
$\calS(t) = e^{-itF_d} \cdots e^{-itF_2} e^{-itF_1}$
associated with
a Hamiltonian $H=\sum_{a=1}^d F_a$.
We assume that
$p,d=O(1)$ are constants independent of $n$, and $F_1,\ldots,F_d\in \calL(k,J)$
for some constants $k$ and $J$ independent of $n$.
Furthermore, we assume that each exponent $e^{-it F_a}$
in the product formula
can be implemented by a constant depth circuit. Accordingly, there exists
a constant $\gamma=O(1)$ independent of $n$ and $t$ such that
$e^{-it \ad_{F_a}}$ maps any single-qubit operator to an operator
supported on at most $\gamma$ qubits.
This setup covers most of  quantum spin Hamiltonians studied in physics
and Trotter-Suzuki product formulas.
\begin{lemma}
Under the above assumptions the coefficients $a_2$ and $a_3$
 in Theorem~\ref{thm:MPF} scale at most linearly with
 the number of qubits $n$.
\end{lemma}
\begin{proof}
Let  $\beta_{p,\ell}(t)$ be the function defined in Eq.~(\ref{beta_comm2}).
Recall that $\ell \in [0,p]$ is an integer.
By definition, each coefficient $a_2$ and $a_3$ is a linear combination
of $O(1)$ terms proportional to  $\beta_{p,\ell}(t)$ with constant coefficients.
Thus  it suffices to prove that  $\beta_{p,\ell}(t)=O(n)$.
Using the definition of $\beta_{p,\ell}(t)$, it suffices to prove that
\be
\label{nested_comm_linear_scaling}
\left\| (\ad_{H})^\ell e^{-i\tau_d \ad_{F_d}}\cdots e^{-i\tau_1 \ad_{F_1}}
(\ad_{F_d})^{q_d}
 \cdots (\ad_{F_{a}})^{q_{a}}(F_{a-1})\right\| = O(n)
\ee
for any evolution times $\tau_i$, integer $a\in [2,d]$, and integers $q_a\ldots,q_d\ge 0$
such that $q_a+\ldots+q_d=p$.
The following proposition controls how  locality and  interaction strength
grow upon taking the commutators.
\begin{prop}
\label{prop:comm}
Suppose $F_1\in \calL(k_1,J_1)$
and $F_2\in \calL(k_2,J_2)$ are arbitrary operators.
Then $[F_1,F_2] \in \calL(k,J)$ with $k=k_1+k_2-1$
and $J=2J_1J_2(k_1+k_2)$.
\end{prop}
\begin{proof}
Let $F_i=\sum_\alpha F_{i,\alpha}$  be a
decompositions of $F_i$ into $k_i$-qubit operators $F_{i,\alpha}$ such that
\[
\sum_{\alpha \, : \, \mathrm{supp}(F_{i,\alpha})\ni q} \; \;
\| F_{i,\alpha}\| \le J_i
\]
for any qubit $q$.
Then
\[
G:=[F_1,F_2] = \sum_{\alpha,\beta} G_{\alpha,\beta}, \qquad G_{\alpha,\beta}=[F_{1,\alpha}, F_{2,\beta}].
\]
Consider any qubit $q$. If $q\in \mathrm{supp}(G_{\alpha,\beta})$
then $q\in \mathrm{supp}(F_{1,\alpha})$ or $q\in \mathrm{supp}(F_{2,\beta})$.
Thus
\be
\label{comm_norm_app}
\sum_{\alpha,\beta \, :\,  \mathrm{supp}(G_{\alpha,\beta})\ni q}\; \;
\| G_{\alpha,\beta}\|
\le \sum_{\alpha \, : \,  \mathrm{supp}(F_{1,\alpha})\ni q}\; \;  \sum_\beta \| G_{\alpha,\beta}\|
+  \sum_{\beta \, : \,  \mathrm{supp}(F_{2,\beta})\ni q}\; \; \sum _\alpha \| G_{\alpha,\beta}\|.
\ee
Clearly $G_{\alpha,\beta}=0$ unless $\mathrm{supp}(F_{1,\alpha})\cap \mathrm{supp}(F_{2,\beta})\ne \emptyset$.
Thus $G_{\alpha,\beta}$ has support on at most $k_1+k_2-1$ qubits,
that is, $G$ has locality $k_1+k_2-1$. For any fixed $\alpha$
one has
\[
\sum_\beta \|G_{\alpha,\beta}\| \le  \sum_{i\in \mathrm{supp}(F_{1,\alpha})}\; \;
\sum_{\beta\, : \, \mathrm{supp}(F_{2,\beta})\ni i}\; \; 2 \|F_{1,\alpha}\| \cdot \|
F_{2,\beta}\| \le 2 J_2 k_1 \|F_{1,\alpha}\|.
 \]
Here we used a bound $\|[F_{1,\alpha},F_{2,\beta}]\|\le 2\|F_{1,\alpha}\|\cdot \|F_{2,\beta}\|$.
Likewise, for any fixed $\beta$ one has
\[
\sum_\alpha \|G_{\alpha,\beta}\| \le  \sum_{i\in \mathrm{supp}(F_{2,\beta})}\; \;
\sum_{\alpha\, : \, \mathrm{supp}(F_{1,\alpha})\ni i}\; \; 2 \|F_{1,\alpha}\| \cdot \|
F_{2,\beta}\| \le 2 J_1 k_2 \|F_{2,\beta}\|.
 \]
Substituting this into Eq.~(\ref{comm_norm_app}) one gets
\[
 \sum_{\alpha,\beta \, :\,  \mathrm{supp}(G_{\alpha,\beta})\ni q}\; \;
\| G_{\alpha,\beta}\|
\le 2J_1 J_2(k_1+k_2).
\]
\end{proof}
Applying Proposition~\ref{prop:comm} recursively $p$ times one infers that
\[
O:=(\ad_{F_d})^{q_d}
 \cdots (\ad_{F_{a}})^{q_{a}}(F_{a-1}) \in \calL(k',J')
\]
where $k'\le k(1+p)$ and $J'\le  J^{1+p} 2^p k(p+1)$.
Here we used  $q_a+\ldots+q_d=p$.
Thus we can write
\[
O = \sum_\alpha O_\alpha
\]
where $O_\alpha$ acts non-trivially on at most $k'$ qubits
and
\[
\sum_{\alpha \, : \, \mathrm{supp}(O_\alpha)\ni q} \; \;
\| O_\alpha\| \le J'
\]
for any qubit $q$.
Define operators
\[
P_\alpha = e^{-i\tau_d \ad_{F_d}}\cdots e^{-i\tau_1 \ad_{F_1}}(O_\alpha).
\]
Unitarity implies $\|P_\alpha\|=\|O_\alpha\|$.
Suppose $P_\alpha$ acts non-trivially on some qubit $q$.
Then $O_\alpha$ acts non-trivially on at least one qubit
in the backward   lightcone of the qubit $q$
with respect to the circuit $e^{-i\tau_d F_d}\cdots e^{-i\tau_1 F_1}$.
Let this lightcone be $L_q$.
The assumption that
$e^{-i\tau \ad_{F_a}}$ maps any single-qubit operator to an operator
supported on at most $\gamma$ qubits implies that
$|L_q|\le \gamma^d$.
Thus
\[
\sum_{\alpha \, : \, \mathrm{supp}(P_\alpha)\ni q} \; \;
\| P_\alpha\| \le \sum_{i\in L_q} \;  \sum_{\alpha \, : \, \mathrm{supp}(O_\alpha)\ni i} \; \; \|O_\alpha\| \le \gamma^d J'.
\]
The same lightcone argument shows that $P_\alpha$ acts non-trivially on at most $k'\gamma^d$ qubits. We conclude that
\be
e^{-i\tau_d \ad_{F_d}}\cdots e^{-i\tau_1 \ad_{F_1}}(O)\in \calL(k'',J''),
\ee
where $k''=\gamma^d k'$ and $J''=\gamma^d J'$.
Finally, we note that $H\in \calL(k,dJ)$. Applying Proposition~\ref{prop:comm}
recursively $\ell$ times one gets
\be
 (\ad_{H})^\ell e^{-i\tau_d \ad_{F_d}}\cdots e^{-i\tau_1 \ad_{F_1}}
(\ad_{F_d})^{q_d}
 \cdots (\ad_{F_{a}})^{q_{a}}(F_{a-1}) \in \calL(k''',J''')
\ee
where $k'''\le k'' + k\ell$ and $J'''\le 2^\ell d^\ell J^\ell J''(\ell k + k'')$.
Combining all above bounds gives
\[
k'''\le \gamma^d kp \quad \mbox{and} \quad
J'''\le 4^{p+2} p^2 d^p J^{2p+1} \gamma^{2d} k^2.
\]
Clearly, $J'''$ is upper bounded by a constant independent of $n$.
It remains to note that $\|X\|\le n J'''$ for any operator $X\in \calL(k''',J''')$.
This proves Eq.~(\ref{nested_comm_linear_scaling}).
 \end{proof}

 \section{Proof of~\Cref{t:mdMPF-error}}\label{sec:minimax_dae}

Before introducing the error bound let $\mathbb{1}$ denote $r$-dimensional vector of ones, let $q_s=2$ for all $s<j$, provided the final time-step $t_j$ is such that $j>1$, and $q_j=1$, finally set $P_0=\bar M^\top(t_0) \bar M(t_0)+\mathbb{1} \mathbb{1}^\top+\varepsilon^2 I$ and
\begin{align*}
     P_s&=\mathbb{1} \mathbb{1}^\top +q_s\varepsilon^2 I+\bar M^\top(t_s) \bar M(t_s)\\
 &-\bar M^\top(t_s) \bar A(t_s)(P_{s-1}+\bar A^\top(t_{s-1})\bar A(t_{s-1}))^{-1}\bar A^\top(t_s)\bar M(t_s)
\end{align*}
Note that $P_j$ is a real symmetric positive-definite $r\times r$-matrix.


\begin{theorem}\label{t:mdMPF-error}
     Assume that~\cref{eq:MQellips} holds and let $\mu^S(t)=\sum_{i=1}^r c_i \rho_{k_i}(t)$ with $c_i$ defined as in Theorem~\ref{thm:MPF}. In the setting of Algorithm~\ref{alg:mdMPF} let \[
    \gamma(t)=
    \left( \sum_{i=1}^r \frac{|c_i|}{k_i^{2p}}\right)
 ( a_1 t^{2p+2} + a_2 t^{2p+1} + a_3 t^{2p} ) + \frac{2\alpha_p (dt)^{p+1}}{(p+1)!k_0^p}
\]
Let $t_j$ and $\hat c(t_j)$ be defined as in Algorithm~\ref{alg:mdMPF} and let $\Psi_j= 2\varepsilon\|\oc(t_j)\|_2+ 4\varepsilon\sum_{s=1}^{j-1}\|\oc(t_s)\|_2$. Then 
for any vector $\ell\in\RR^r$ we have:
\begin{equation}
  \label{eq:mnmx-err2}
   |\ell^\top(\oc(t_j)-\hat c(t_j)|\le 2\hat\beta^\frac12_j \langle P^+_j\ell,\ell\rangle^\frac 12
\end{equation}
    Here $\hat\beta_j:=\left(\sum_{s\le j-1} \sqrt{r} \gamma(t_s)+\Psi_j\right)^2-\alpha_j+ \langle P_j\hat c(t_j),\hat c(t_j)\rangle{}$ with $r_0=\mathbb{1}+\oc(t_0)$, $\alpha_0=1+\oc(t_0)^\top \oc(t_0)$ and
\begin{equation*}
    \begin{split}
      &\alpha_j=\alpha_{j-1}+1-r_{j-1}^\top (P_{j-1}+\bar A^\top(t_{j-1})\bar A(t_{j-1}))^+ r_{j-1}, \\
    &r_j=\bar M^\top(t_j) \bar A(t_{j-1})(P_{j-1}+\bar A^\top(t_{j-1})\bar A(t_{j-1}))^+r_{j-1}+\mathbb{1},
          \end{split}
        \end{equation*}
\end{theorem}

To prove~\Cref{t:mdMPF-error} we need to recall basic results from minimax estimation for linear discrete-time differential-algebraic equations~\cite{zhuk2010minimax}, namely the notion of the minimax estimate. Assume that we are given a vector $y$ in the form of $y=C\varphi+\eta$ where $C$ is a given linear transformation of a vector $\varphi$ from an abstract Hilbert space $\calH$ with inner product $\langle\cdot,\cdot\rangle$. We further assume that $\varphi$ solves a linear equation $L\varphi=f$ for a given bounded transformation $L$, and that $f$ and $\eta$ are unknown elements satisfying $$
G=\{(f,\eta):\langle Q_1f,f\rangle+\langle Q_2\eta,\eta\rangle\le 1\}
$$ for given symmetric positive definite operators $Q_{1,2}$. The vector $\hat\varphi \in G_y=\{\varphi: \la Q_1L\varphi,L\varphi\ra+\la Q_2(y-C\varphi),y-C\varphi\ra\le 1\}$ is called a minimax estimate of $\varphi$ if
\be\label{eq:minimax-estimate}
\forall\ell\in\calH: \quad \max_{\psi\in G_y}|\langle \ell,\hat\varphi-\psi\rangle| = \min_{\varphi\in G_y}\max_{\psi\in G_y}|\langle \ell,\varphi-\psi\rangle|
\ee
and $\hat\rho(\ell):=\max_{\psi\in G_y}|\langle \ell,\hat\varphi-\psi\rangle|$ is called the minimax error. Clearly, $\hat\varphi$ is the minimax or Tchebyshev center of the set of all admissible $\varphi\in G_y$ which are compatible with the uncertainty description given by $G$.
\begin{theorem}[Theorem 1 from~\cite{zhuk2010minimax}]\label{p:gnrl}
Let \(
I(\varphi):=\la Q_1L \varphi,L\varphi\ra+\la Q_2(y-C\varphi),y-C\varphi\ra
\).
Then the minimax estimate $\hat\varphi$ is the unique minimizer of $I$: $\hat\varphi\in\mathrm{Argmin}_\phi I(\phi)$, and the minimax error is
\begin{align*}
    \hat\rho(\ell)=
(1-I(\hat \varphi))^\frac 12 \max_{\phi: \la (L^\dag Q_1 L + C^\dag Q_2 C )\phi,\phi\ra\le 1}\la\ell,\phi\ra
=(1-I(\hat \varphi))^\frac 12 \la (L^\dag Q_1 L + C^\dag Q_2 C )^{-1}\ell,\ell\ra^\frac 12
\end{align*}
provided $L^\dag Q_1 L + H^\dag Q_2 H $ is invertible. Moreover
\begin{equation}
 \label{eq:c:vecest}
 \begin{split}
    & \max_{\phi\in G_y}\|\hat \varphi-\phi\|=
    \min_{\varphi\in G_y}\max_{\phi\in G_y}\|\varphi-\phi\|=
    (1-I(\hat \varphi))^\frac 12
    \max_{\|\ell\|=1} \la (L^\dag Q_1 L + C^\dag Q_2 C )^{-1}\ell,\ell\ra^\frac 12
  \end{split}
\end{equation}
\end{theorem}
Now, let us prove~\Cref{t:mdMPF-error}.
\begin{proof}[Proof of~\Cref{t:mdMPF-error}]
  Recall the setting of Algorithm~\ref{alg:mdMPF}. The idea of this proof is to construct a set $\calG_y$ of all feasible discrete-time trajectories $\varphi=(\vc(t_0)\dots\vc(t_j))^\top$, i.e. all trajectories which satisfy~\eqref{eq:stateq} for all possible realizations of $E^D_{1,2}(t_0)\dots E^D_{1,2}(t_j)$, and then demonstrate that the discrete-time trajectory of exact coefficients, $\varphi^\star=(\oc(t_0)\dots\oc(t_j))^\top$ belongs to $\calG_y$ as well as the trajectory $\hat\varphi=(\hat c(t_0)\dots \hat c(t_j))^\top$ generated by Algorithm~\ref{alg:mdMPF}. Then the worst-case error of approximating individual components of $\varphi^\star$ by corresponding com[onents of $\hat\varphi$ as per~\eqref{eq:minimax-estimate} will be given by~\eqref{eq:c:vecest}.

  Without loss of generality assume that $\oc(t_0)$ is known. Recall that by assumption the exact matrices $M,Q$ admit the following representation: $M(t)\in\{\bar M(t)+E^D_{1}(t), \|E^D_{1}(t)\|_{2,2}\le \varepsilon\}$ and $Q(t)\in\{\bar A(t) + E^D_2(t), \|E^D_{2}(t)\|_{2,2}\le \varepsilon\}$, and there exist $E^\star_{1,2}$ such that $M(t_j)=\bar M(t_j)+E^\star_1(t_j)$ and $Q(t_j)=\bar A(t_j)+E^\star_2(t_j)$ for some unknown $E^\star_{1,2}$, see~\eqref{eq:MQellips} above. Then, adding and substructing $E^D_{1,2}$ in~\eqref{eq:stateq} we find that $\oc$ satisfies the following equation:
  \begin{align}
    \label{eq:stateq1}
  (\bar M(t_j) + E^D_1) \vc(t_j) &= (\bar A(t_j)+ E^D_2) \vc(t_{j-1}) + L^{\exact}(t_j) -  L^{\apx}(t_{j}) + (E^D_1-E_1^\star)\oc(t_j)+ (E_2^\star - E^D_2)\oc(t_{j-1})\\
  &=(\bar A(t_j)+ E^D_2) \vc(t_{j-1}) + F_j, \quad F_j=(\bar M(t_j)+E^D_1)\oc(t_j)-(\bar A(t_j)+E^D_2)\oc(t_{j-1})
\end{align}
We stress that since $\oc, E^D_{1,2}$ are not known the term $F_j$ is treated as an unknown input. The purpose of introducing~\eqref{eq:stateq1} is to demonstrate that there exist input $F_j$ such that $\oc(t_j)$ satisfies~\eqref{eq:stateq1}, namely when we set $\vc=\oc$ the equation holds~\eqref{eq:stateq1}, hence $\oc$ is contained in the set of all $\vc$ satisfying~\eqref{eq:stateq1} while $F_j$ runs through a set $\{F:F=(\bar M(t_j)+E^D_1)\oc(t_j)-(\bar A(t_j)+E^D_2)\oc(t_{j-1}), \|E^D_{1,2}\|_{2,2}\le\varepsilon\}$. Moreover, it is obvious that $\|(\bar M(t_j) + E^D_1) \vc(t_j)-(\bar A(t_j)+ E^D_2) \vc(t_{j-1})\|_2=\|F_j\|_2$ hence the discrete-time trajectory $\varphi^\star=(\oc(t_0)\dots\oc(t_j))^\top$ belongs to the set of all trajectories $\varphi =(\vc(t_0)\dots\vc(t_j))^\top$ such that:
  \begin{align}\label{eq:stateq2}
    \|\vc(t_0)-\oc(t_0)\|_2+\sum_{s=1}^j \|(\bar M(t_s) + E^D_1(t_s)) \vc(t_s)-(\bar A(t_s)+ E^D_2(t_s)) \vc(t_{s-1})\|_2
    \le \sum_{s=1}^j \|F_j\|_2,\quad \mathbb{1}^\top \vc(t_s)=1
  \end{align}
  Using sub-multiplicativity of the matrix norm $\|\cdot\|_{2,2}$ we find:
  \begin{align*}
    \|(\bar M(t_j) + E^D_1) \vc(t_j)-(\bar A(t_j)+ E^D_2) \vc(t_{j-1})\|_2  &\le  \|\bar M(t_j) \vc(t_j)-\bar A(t_j)\vc(t_{j-1})\|_2+\|E^D_1 \|_{2,2}\|\vc(t_s)\|_2+ \|E^D_2\|_{2,2}\|\vc(t_{s-1})\|_2\\
    &\le \|\bar M(t_j) \vc(t_j)-\bar A(t_j)\vc(t_{j-1})\|_2 + \varepsilon \| \vc(t_s)\|_2+\varepsilon \|\vc(t_{s-1})\|_2
  \end{align*}
  and this upper-bound is attained at
\[
  \bar E_1^D(t_s)=\varepsilon \frac{\bigl[\bar M(t_s)\vc(t_s)-(\bar A(t_s)+ E^D_2(t_s)) \vc(t_{s-1})\bigr]\vc^\top(t_s)}{\|\bar M(t_s)\vc(t_s)-(\bar A(t_s)+ E^D_2(t_s)) \vc(t_{s-1})\|_2\|\vc(t_s)\|_2}, \quad \bar E_2^D(t_s)=\varepsilon \frac{\bigl[\bar M(t_s)\vc(t_s)-\bar A(t_s)\vc(t_{s-1})\bigr]\vc^\top(t_{s-1})}{\|\bar M(t_s)\vc(t_s)-\bar A(t_s)\vc(t_{s-1})\|_2\|\vc(t_{s-1})\|_2}\]
Indeed, simple algebra shows that $\|\bar E_{1,2}^D\|_{2,2}=\varepsilon$ and \[
\|(\bar M(t_j) + \bar E^D_1) \vc(t_j)-(\bar A(t_j)+ \bar E^D_2) \vc(t_{j-1})\|_2  = \|\bar M(t_j) \vc(t_j)-\bar A(t_j)\vc(t_{j-1})\|_2+\varepsilon \|\vc(t_s)\|_2+\varepsilon \|\vc(t_{s-1})\|_2
\]
In other words the worst-case realization of the matrices $E_{1,2}^D$ are given by the trajectories $\bar E_{1,2}^D(t_s)$, and for this worst-case scenario~\eqref{eq:stateq2} takes the following form:
\begin{align}\label{eq:stateq3}
\|\vc(t_0)-\oc(t_0)\|_2+\sum_{s=1}^j \bigl(\|\bar M(t_s)\vc(t_s)-\bar A(t_s)\vc(t_{s-1})\|_2 +\varepsilon \|\vc(t_s)\|_2+\varepsilon \|\vc(t_{s-1})\|_2\bigr)\le \sum_{s=1}^j \|\bar F_j\|_2, \quad \mathbb{1}^\top\vc(t_s)=1
\end{align}
provided $\bar F_j = (\bar M+\bar E^D_1)\oc(t_j)-(\bar A+\bar E^D_2)\oc(t_{j-1})$. Setting $\vc=\oc$ in the definition of $\bar E_{1,2}^D(t_s)$ we find:
\[
\bar F_j = \|\bar M(t_s)\oc(t_s)-\bar A(t_s)\oc(t_{s-1})\|_2 +\varepsilon \|\oc(t_s)\|_2+\varepsilon \|\oc(t_{s-1})\|_2
\] hence we can further restrict the set~\eqref{eq:stateq3} of all feasible trajectories $\varphi$ to the following set:
\begin{align}\label{eq:stateq4}
  \calG=\{\varphi:\|\vc_0-\oc(t_0)\|_2&+\sum_{s=1}^j \bigl(\|\bar M(t_s)\vc(t_s)-\bar A(t_s)\vc(t_{s-1})\|_2 +\varepsilon \|\vc(t_s)\|_2+\varepsilon \|\vc(t_{s-1})\|_2\bigr)
                        \le \Sigma_j,  \mathbb{1}^\top\vc(t_s)=1,\\
  &\Sigma_j=\sum_{s=1}^j\bigl(\|\bar M(t_s)\oc(t_s)-\bar A(t_s)\oc(t_{s-1})\|_2 +\varepsilon \|\oc(t_s)\|_2+\varepsilon \|\oc(t_{s-1})\|_2\bigr)\}
\end{align}
Clearly, $\varphi^\star\in\calG$: indeed, substituting $\varphi=\varphi^\star$ the inequality in the 1st line of the definition of $\calG$ becomes exact. Note that $\hat\varphi\in\calG$: indeed, by construction of $\hat\varphi$ the inequality in the 1st line of the definition of $\calG$ is strict. Finally, since $\|\cdot\|_2\le\|\cdot\|_1$ it follows that
\begin{align*}
\|\vc(t_0)-\oc(t_0)\|^2_2&+\sum_{s=1}^j \bigl(\|\bar M(t_s)\vc(t_s)-\bar A(t_s)\vc(t_{s-1})\|^2_2 +\varepsilon^2 \|\vc(t_s)\|^2_2+\varepsilon^2 \|\vc(t_{s-1})\|_2^2\bigr)\\
 &\le \left(\|\vc_0-\oc(t_0)\|_2+\sum_{s=1}^j \bigl(\|(\bar M(t_s)\vc(t_s)-\bar A(t_s)\vc(t_{s-1})\|_2 +\varepsilon\|\vc(t_s)\|_2+\varepsilon \|\vc(t_{s-1})\|_2\bigr)\right)^2\le \Sigma_j^2
\end{align*}
hence both $\varphi^\star=(\oc(t_0)\dots \oc(t_j))^\top$ and $\hat\varphi=(\vc(t_0)\dots \vc(t_j))^\top$ belong to the set:
\begin{align}
  \label{eq:Gy1}
  \tilde\calG_y=\{\varphi&=(\vc(t_0)\dots\vc(t_j))^\top: \|\vc(t_0)-\oc(t_0)\|^2_2+\varepsilon^2\|\vc(t_0)\|^2_2+\sum_{s=1}^j \|\bar M(t_s)\vc(t_s)-\bar A(t_s)\vc(t_{s-1})\|^2_2\\
  &+\varepsilon^2 \|\vc(t_j)\|_2^2+2\varepsilon^2 \sum_{s=1}^{j-1}\|\vc(t_s)\|^2_2\le \Sigma_j^2,\quad \mathbb{1}^\top\vc(t_s)=1\}
\end{align}
Define $C(\alpha) = \bigl(
\begin{smallmatrix}
  \mathbb{1}^\top\\\alpha I
\end{smallmatrix}
\bigr)$ and $\tilde y = (1,0,\dots,0)^\top$ so that $\|\tilde y - C(\alpha)\vc(t_s)\|_2^2 = (1-\mathbb{1}^\top\vc(t_s))^2+\alpha^2 \|\vc(t_s)\|^2_2$. Using this notations we can rewrite~\eqref{eq:Gy1} in a more compact form: let $\calC_j=\operatorname{diag}(C(\varepsilon^2),C(\sqrt{2}\varepsilon^2),\dots,C(\sqrt{2}\varepsilon^2),C(\varepsilon^2))$, $y_j=(\tilde y\dots,\tilde y)^\top$, $b_j=( \oc(t_0),0,\dots 0)^\top$ and set
\be
L_j=\left[\begin{smallmatrix}
    I&&0&&0&&\hdots&&0&&0\\
    -\bar A(t_0)&& \bar M(t_0)&&0&&\hdots&&0&&0 \\
    0   &&-\bar A(t_1) &&\bar M(t_1)  &&\hdots &&0&&0\\
    \vdots&&\vdots&&\vdots&& \vdots&&\vdots&&\vdots\\
    0&&0&&0&&\hdots&&-\bar A(t_j)&&\bar M(t_j)
  \end{smallmatrix}\right]
\ee
Clearly, every element of~\eqref{eq:Gy1} belongs to the following set $\calG_y$, which is in fact just a compact way of writing~\eqref{eq:Gy1} but with the hard constraint $\mathbb{1}^\top\vc(t_s)=1$ relaxed as $(1-\mathbb{1}^\top\vc(t_s))^2$:
\begin{equation}
  \label{eq:Gy}
  \calG_y=\{\varphi: \|L_j\varphi - b_j\|_2^2 + \|y-\calC_j \varphi\|_2^2\le\Sigma_j^2\}
\end{equation}
Now, as we saw above both $\varphi^\star=(\oc(t_0)\dots \oc(t_j))^\top$ and $\hat\varphi=(\vc(t_0)\dots \vc(t_j))^\top$ belong to $\calG_y$ hence the worst-case error of this trajectory is given by the largest diameter of $\calG_y$. To compute the latter let us first note that by~\eqref{eq:stateq}: \[
\|(\bar M(t_s)\pm E_1^\star)\oc(t_s)-(\bar A(t_s)\pm E_2^\star)\oc(t_{s-1})\|_2\le
\| L^{\exact}(t_s) - L^\apx(t_s)\|_2 + \varepsilon\|\oc(t_s)\|_2+\varepsilon\|\oc(t_{s-1})\|_2
\]
hence
\[
\Sigma_j\le 2\varepsilon\|\oc(t_j)\|_2+\sum_{s=1}^j \| L^{\exact}(t_j) - L^\apx(t_j)\|_2 + 4\varepsilon\sum_{s=1}^{j-1}\|\oc(t_s)\|_2
\] and let's estimate $\| L^{\exact}(t_j) - L^\apx(t_j)\|_2$. To this end
recall that $\rho^{\apx}(t_j)=\calS(dt/k_0)^{k_0} \mu^D(t_{j-1})\calS(dt/k_0)^{-k_0}$ and $\mu^D(t)=\sum_{i=1}^r \oc_i(t)\rho_{k_i}(t)$. Noting that $\|\rho_{k_i}(t_j)\|_\infty = 1$ and $\|\rho_{k_i}(t_j)\|_\infty =1$ we have:
\begin{align*}
    |L_i^{\exact}(t_j) &- L_i^\apx(t_j)| = |\tr\bigl((\rho(t_j)-\rho^{\apx}(t_j))\rho_{k_i}(t_j)\bigr)|
   = |\tr\bigl((e^{-iHdt}\rho(t_{j-1})e^{iHdt}-\rho^{\apx}(t_j))\rho_{k_i}(t_j)\bigr)|\\
   &\le |\tr\bigl((e^{-iHdt}(\rho(t_{j-1})-\mu^D(t_{j-1}))e^{iHdt}\rho_{k_i}(t_j)\bigr)|
   + |\tr\bigl((e^{-iHdt}\mu^D(t_{j-1})e^{iHdt}-\rho^{\apx}(t_j))\rho_{k_i}(t_j)\bigr)|\\
   &\le \|e^{-iHdt}(\rho(t_{j-1})-\mu^D(t_{j-1}))e^{iHdt}\|_2 \|\rho_{k_i}(t_j)\|_2+\|e^{-iHdt}\mu^D(t_{j-1})e^{iHdt}-\rho^{\apx}(t_j)\|_1 \|\rho_{k_i}(t_j)\|_\infty \\
   &\overset{by~\eqref{childs_su}}{\le} \|\rho(t_{k-j})-\mu^D(t_{j-1})\|_2  + \frac{2\alpha_p (dt)^{p+1}}{(p+1)!k_0^p})\\
   &\le \|\rho(t_{j-1})-\mu^S(t_{j-1})\|_2 + \frac{2\alpha_p (dt)^{p+1}}{(p+1)!k_0^p})\\
   &\le \gamma(t_{j-1})\Rightarrow \| L^{\exact}(t_j) - L^\apx(t_j)\|_2 \le \sqrt{r}\,\gamma(t_{j-1})
\end{align*}
Hence for $\Psi_j$ defined as in theorem's statement we have: \[
\Sigma_j\le \Psi_j + \sqrt{r}\sum_{s\le j-1} \gamma(t_{s})
\]
Now, to complete the proof we compute the largest diameter of $\calG_y$: it is sufficient to set $Q_1=Q_2=\Sigma^{-2}$ and $L=L_j$, $C=C_j$ and apply~\eqref{eq:c:vecest} in order to compute the radius of $\calG_y$ which if doubled gives error representation in the computational form using matrix $P_j$ as in the r.h.s. of \eqref{eq:mnmx-err2}. Detailed derivation of how to compute~\eqref{eq:c:vecest} recursively using $P_j$ is given in~\cite{zhuk2010minimax}.
\end{proof}

\bibliographystyle{unsrt}
\bibliography{mybib}

\begin{thebibliography}{10}

\bibitem{wang2008quantum}
Hefeng Wang, Sabre Kais, Al{\'a}n Aspuru-Guzik, and Mark~R Hoffmann.
\newblock Quantum algorithm for obtaining the energy spectrum of molecular
  systems.
\newblock {\em Physical Chemistry Chemical Physics}, 10(35):5388--5393, 2008.

\bibitem{reiher2017elucidating}
Markus Reiher, Nathan Wiebe, Krysta~M Svore, Dave Wecker, and Matthias Troyer.
\newblock Elucidating reaction mechanisms on quantum computers.
\newblock {\em Proceedings of the national academy of sciences},
  114(29):7555--7560, 2017.

\bibitem{harle2022observing}
Nikhil Harle, Oles Shtanko, and Ramis Movassagh.
\newblock Observing and braiding topological {M}ajorana modes on programmable
  quantum simulators.
\newblock {\em arXiv preprint arXiv:2203.15083}, 2022.

\bibitem{kaufman2016quantum}
Adam~M Kaufman, M~Eric Tai, Alexander Lukin, Matthew Rispoli, Robert Schittko,
  Philipp~M Preiss, and Markus Greiner.
\newblock Quantum thermalization through entanglement in an isolated many-body
  system.
\newblock {\em Science}, 353(6301):794--800, 2016.

\bibitem{garttner2017measuring}
Martin G{\"a}rttner, Justin~G Bohnet, Arghavan Safavi-Naini, Michael~L Wall,
  John~J Bollinger, and Ana~Maria Rey.
\newblock Measuring out-of-time-order correlations and multiple quantum spectra
  in a trapped-ion quantum magnet.
\newblock {\em Nature Physics}, 13(8):781--786, 2017.

\bibitem{lloyd1996universal}
Seth Lloyd.
\newblock Universal quantum simulators.
\newblock {\em Science}, 273(5278):1073--1078, 1996.

\bibitem{low2017optimal}
Guang~Hao Low and Isaac~L Chuang.
\newblock Optimal {H}amiltonian simulation by quantum signal processing.
\newblock {\em Physical review letters}, 118(1):010501, 2017.

\bibitem{childs2021theory}
Andrew~M Childs, Yuan Su, Minh~C Tran, Nathan Wiebe, and Shuchen Zhu.
\newblock Theory of {T}rotter error with commutator scaling.
\newblock {\em Physical Review X}, 11(1):011020, 2021.

\bibitem{temme2017error}
Kristan Temme, Sergey Bravyi, and Jay~M Gambetta.
\newblock Error mitigation for short-depth quantum circuits.
\newblock {\em Physical review letters}, 119(18):180509, 2017.

\bibitem{berg2022probabilistic}
Ewout van~den Berg, Zlatko~K Minev, Abhinav Kandala, and Kristan Temme.
\newblock Probabilistic error cancellation with sparse {P}auli-{L}indblad
  models on noisy quantum processors.
\newblock {\em arXiv preprint arXiv:2201.09866}, 2022.

\bibitem{childs2018toward}
Andrew~M Childs, Dmitri Maslov, Yunseong Nam, Neil~J Ross, and Yuan Su.
\newblock Toward the first quantum simulation with quantum speedup.
\newblock {\em Proceedings of the National Academy of Sciences},
  115(38):9456--9461, 2018.

\bibitem{childs2012hamiltonian}
Andrew~M Childs and Nathan Wiebe.
\newblock Hamiltonian simulation using linear combinations of unitary
  operations.
\newblock {\em arXiv preprint arXiv:1202.5822}, 2012.

\bibitem{low2019well}
Guang~Hao Low, Vadym Kliuchnikov, and Nathan Wiebe.
\newblock Well-conditioned multiproduct {H}amiltonian simulation.
\newblock {\em arXiv preprint arXiv:1907.11679}, 2019.

\bibitem{vazquez2022well}
Almudena~Carrera Vazquez, Daniel~J Egger, David Ochsner, and Stefan Woerner.
\newblock Well-conditioned multi-product formulas for hardware-friendly
  hamiltonian simulation.
\newblock {\em arXiv preprint arXiv:2207.11268}, 2022.

\bibitem{rendon2022improved}
Gumaro Rendon, Jacob Watkins, and Nathan Wiebe.
\newblock Improved error scaling for trotter simulations through extrapolation.
\newblock {\em arXiv preprint arXiv:2212.14144}, 2022.

\bibitem{endo2019mitigating}
Suguru Endo, Qi~Zhao, Ying Li, Simon Benjamin, and Xiao Yuan.
\newblock Mitigating algorithmic errors in a hamiltonian simulation.
\newblock {\em Physical Review A}, 99(1):012334, 2019.

\bibitem{chin2010multi}
Siu~A Chin.
\newblock Multi-product splitting and {R}unge-{K}utta-{N}ystr{\"o}m
  integrators.
\newblock {\em Celestial Mechanics and Dynamical Astronomy}, 106:391--406,
  2010.

\bibitem{zhuk2010minimax}
Sergiy~M Zhuk.
\newblock Minimax state estimation for linear discrete-time
  differential-algebraic equations.
\newblock {\em Automatica}, 46(11):1785--1789, 2010.

\bibitem{zhuk2021minimax}
Sergiy Zhuk, Orest~V Iftime, Jonathan~P Epperlein, and Andrey Polyakov.
\newblock Minimax sliding mode control design for linear evolution equations
  with noisy measurements and uncertain inputs.
\newblock {\em Systems \& Control Letters}, 147:104830, 2021.

\bibitem{zhuk2023detectability}
Sergiy Zhuk, Mykhaylo Zayats, and Emilia Fridman.
\newblock Detectability and global observer design for 2d navier--stokes
  equations with uncertain inputs.
\newblock {\em Automatica}, 153:111043, 2023.

\bibitem{nielsen2002quantum}
Michael~A Nielsen and Isaac Chuang.
\newblock Quantum computation and quantum information, 2002.

\bibitem{kim2023scalable}
Youngseok Kim, Christopher~J Wood, Theodore~J Yoder, Seth~T Merkel, Jay~M
  Gambetta, Kristan Temme, and Abhinav Kandala.
\newblock Scalable error mitigation for noisy quantum circuits produces
  competitive expectation values.
\newblock {\em Nature Physics}, pages 1--8, 2023.

\bibitem{diamond2016cvxpy}
Steven Diamond and Stephen Boyd.
\newblock {CVXPY}: {A} {P}ython-embedded modeling language for convex
  optimization.
\newblock {\em Journal of Machine Learning Research}, 17(83):1--5, 2016.

\bibitem{andrews_askey_roy_1999}
George~E. Andrews, Richard Askey, and Ranjan Roy.
\newblock {\em Special Functions}.
\newblock Encyclopedia of Mathematics and its Applications. Cambridge
  University Press, 1999.

\bibitem{kitaev2002classical}
Alexei~Yu Kitaev, Alexander Shen, and Mikhail~N Vyalyi.
\newblock {\em Classical and quantum computation}.
\newblock Number~47. American Mathematical Soc., 2002.

\end{thebibliography}

\end{document}